\documentclass[aps,article,citeautoscript,footinbib,superscriptaddress,prx]{revtex4-2}
\usepackage[T1]{fontenc}
\usepackage[utf8]{inputenc}
\usepackage{lipsum}
\usepackage{tcolorbox}
\usepackage{changepage}
\usepackage{amsmath,bm}
\usepackage{amssymb}
\usepackage{yfonts}
\usepackage{eufrak}
\usepackage{bbm}
\usepackage{braket}
\usepackage{graphicx}
\usepackage{xcolor}
\usepackage{pifont}
\usepackage[mathscr]{euscript}
\usepackage[shortlabels]{enumitem}
\usepackage[justification=justified]{subcaption}
\usepackage{float}
\usepackage{dsfont}
\usepackage{comment}
\usepackage{subcaption}
\usepackage{slashed}
\usepackage[normalem]{ulem} 
\usepackage{bbm} 
\usepackage[section]{placeins}
\usepackage{tikz-feynman}
\usepackage[papersize={8.5in,11in}]{geometry}
\usepackage[colorlinks=true]{hyperref}  
\hypersetup{
    unicode=false,          
    pdftoolbar=true,        
    pdfmenubar=true,        
    pdffitwindow=false,     
    pdfstartview={FitH},    
    pdftitle={Flstar},    
    pdfauthor={Bonetti, Christos, Sachdev, Pandey, Shanker, Sharma},     
    pdfsubject={},   
    pdfcreator={},   
    pdfproducer={}, 
    pdfkeywords={} {} {}, 
    pdfnewwindow=true,      
    colorlinks=true,       
    linkcolor=blue, 
    citecolor=blue,        
    filecolor=magenta,      
    urlcolor=blue           
} 

\usepackage{xcolor}
\definecolor{brandeisblue}{rgb}{0.0, 0.44, 1.0}

\usepackage{mathtools} 

\newcommand{\vi}{{\boldsymbol{i}}}
\newcommand{\vj}{{\boldsymbol{j}}}
\newcommand{\bk}{\mathbf{k}}
\newcommand{\bq}{\mathbf{q}}

\begin{document}
\begin{flushright}
 \href{https://arxiv.org/abs/2507.05336}{arXiv:2507.05336} 
\end{flushright}


\title{Thermal SU(2) lattice gauge theory for intertwined orders\\ and hole pockets in the cuprates}

\begin{abstract} 
The cuprate pseudogap phase displays Fermi arc spectral weight in photoemission and scanning tunneling microscopy (STM), while recent magnetotransport observations yield evidence for the existence of hole pockets of fractional area $p/8$, where $p$ is the doping density. We present
a Monte Carlo study of a thermal SU(2) lattice gauge theory which can reconcile these observations. Our simulation includes the SU(2) gauge field $U$ of a $\pi$-flux spin liquid, and a SU(2) fundamental charge $e$ Higgs boson $B$. There is a Yukawa coupling between $B$, the fermionic spinons of the spin liquid, and the hole pockets of a fractionalized Fermi liquid.
At the higher temperatures of the pseudogap, the finite-doping sign problem is evaded by including only thermal fluctuations of $B$ and $U$, while the fermions are diagonalized exactly for each boson background. Our study also yields a fractionalized description of intertwined orders at lower temperatures, including the onset of $d$-wave superconductivity by the expulsion of vortices with flux $h/(2e)$, each with charge-order halos. We discuss conditions under which quantum oscillations in the density of states from hole pockets of area $p/8$ could be observable in clean under-hole-doped cuprates.
\end{abstract}

\author{Harshit Pandey}
\affiliation{The Institute of Mathematical Sciences, Chennai 600113, India}
\affiliation{Homi Bhabha National Institute, Training School Complex, Anushaktinagar, Mumbai 400094, India}

\author{Maine Christos}
\affiliation{Department of Physics, Harvard University, Cambridge MA 02138, USA}
\affiliation{Department of Physics and Institute for Quantum Information and Matter,
California Institute of Technology, Pasadena, CA 91125, USA}

\author{Pietro M. Bonetti}
\affiliation{Department of Physics, Harvard University, Cambridge MA 02138, USA}

\author{Ravi Shanker}
\affiliation{The Institute of Mathematical Sciences, Chennai 600113, India}
\affiliation{Homi Bhabha National Institute, Training School Complex, Anushaktinagar, Mumbai 400094, India}

\author{Sayantan Sharma}
\affiliation{The Institute of Mathematical Sciences, Chennai 600113, India}
\affiliation{Homi Bhabha National Institute, Training School Complex, Anushaktinagar, Mumbai 400094, India}

\author{Subir Sachdev}
\affiliation{Department of Physics, Harvard University, Cambridge MA 02138, USA}
\affiliation{Center for Computational Quantum Physics, Flatiron Institute, 162 5th Avenue, New York, NY 10010, USA}
\affiliation{The Abdus Salam International Centre for Theoretical Physics, Strada Costiera 11, I-34151, Trieste, Italy.}

\maketitle

\begin{adjustwidth}{1.5cm}{1.5cm}
\begin{tcolorbox}
\noindent
{\bf Significance Statement:} The hole-doped cuprate superconductors have the highest critical temperatures under ambient pressure among all known materials. These materials are also unique in having a `pseudogap' metal phase above the critical temperature, suggesting a connection of this phase to the high critical temperatures. Some recent experiments support a  theory for the pseudogap phase which has an entangled quantum spin liquid co-existing with ordinary electronic quasiparticles with their momenta in small pockets. Our work performs Monte Carlo simulations on a lattice gauge model of this theory to also understand experiments in which light ejects electrons from the sample. We also give a perspective on intertwined low temperature ordering phenomena in the lattice gauge model.
\end{tcolorbox} 
\end{adjustwidth}

\newpage
\linespread{1.1}
\tableofcontents

\section{Introduction}
\label{sec:intro}

The hole-doped copper oxide (`cuprate’) superconductors exhibit the highest known superconducting critical temperatures ($T_c$) at ambient pressure. A distinctive feature of these materials is the presence of a pseudogap phase at temperatures ($T$) above $T_c$ at low hole doping levels, $p$, above those with  antiferromagnetic order. This  feature suggests a causal relationship between the pseudogap and high-$T_c$ superconductivity. Consequently, elucidating the structure of the pseudogap and the nature of its transition to $d$-wave superconductivity remains a central challenge in the theory of quantum matter.

The electronic excitation spectrum of the overdoped cuprates at large $p$, beyond the pseudogap phase, has a `large' Fermi surface of zero energy excitations \cite{Dama05,Hussey08}.  This large Fermi surface encloses the conventional Luttinger area of $(1+p)/2$ (the factor of 2 is from the spin degeneracy), expressed as a fraction of the area of the square lattice Brillouin zone. In contrast, in the pseudogap phase at small $p$, for probes that eject electrons from the sample, the Fermi surface is truncated 
by an energy gap in the `anti-nodal' region of the Brillouin zone on a square lattice, near momenta $(\pi, 0)$, $(0,\pi)$ \cite{Shen11}. This leaves the hallmark `Fermi arcs' across the Brillouin zone diagonals in the pseudogap phase, observed in photoemission \cite{Norman98,ShenShen05,Johnson11,Kondo20,Kondo23,Damascelli25} and scanning tunneling microscopy (STM) \cite{Hoffman14,Davis14}
at dopings without antiferromagnetic order at $T=0$. 

On the other hand, recent magnetotransport experiments \cite{Ramshaw20,Yamaji24}, which do not eject electrons from the sample, paint a rather different picture of the pseudogap quasiparticle spectrum. These provide compelling evidence for the existence of hole pockets across the Brillouin zone diagonals, with quasiparticles which can tunnel coherently between square lattice layers.
Notably, the Yamaji effect measurements by Chan {\it et al.} \cite{Yamaji24} in HgBa$_2$CuO$_{4+\delta}$ determine the 
hole pocket area based on the $c$-axis lattice spacing and the observed Yamaji angle, yielding a fractional area 
of approximately $1.3\%$ at doping $p = 0.1$.

One influential perspective---the phase fluctuation framework of Emery and Kivelson \cite{EmeryKivelson}—models the pseudogap as a fluctuating superconducting state. In a Born-Oppenheimer approach, a classical XY model captures the thermal fluctuations of the superconducting phase, which modulates the pairing amplitude of the Bogoliubov Hamiltonian for quantum electrons. This approach has successfully explained a range of experimental observations \cite{Franz98,Scalapino02,Dagotto05,Berg07,Li_2010,Li11JPhys,Li11PRB,Li_2011,Sumilan17,Majumdar22,YQi23,Xiang24,yang2025}, including the Fermi arc features.
But the phase fluctuation picture is difficult to reconcile with the pockets detected in magnetotransport.

Another perspective is to examine a fluctuating spin density wave (SDW), {\it i.e.\/} antiferromagnetic, order parameter (which is a vector in spin space, and hence carries spin $S=1$) in the background of a Fermi liquid with a conventional 
Luttinger-volume large Fermi surface \cite{SchmalianPines1,SchmalianPines2,Tremblay04,Shen11,Chubukov23,Chubukov25}. Here too, the focus is on classical thermal fluctuations of the order parameter. This approach yields a convincing theory in regimes where the ground state has long-range SDW order and the large Fermi surface has been reconstructed into pockets. Such a theory has been successfully applied to the electron-doped cuprates \cite{Tremblay04,Chubukov25}. However, there are difficulties in applying the SDW fluctuation theory to the hole-doped cuprates beyond the doping where there is no SDW order at $T=0$ \cite{Chubukov23}. 

An alternative class of theories interprets the pseudogap phase as a quantum phase in its own right, and not directly associated with a thermally fluctuating order parameter \cite{Lee89,SS94,LeeWen96,LeeNagaosaWen,YRZ,YRZreview,TSSSMV03,TSSSMV04,APAV04,Bonderson16,ElseSenthil,Stanescu_2006,Kaul07,Georges08,ACL08,Imada09,Qi10,Moon11,Punk12,Punk15,Topo_PNAS,Fabrizio22,ZhangSachdev_ancilla,ZhangSachdev_ancilla2,Tremblay21,Mascot22,Tremblay22,Georges_pseudogap_24,Joshi23}. These theories interpret the `Fermi arcs’ as segments of hole pocket Fermi surfaces, with the back sides exhibiting suppressed spectral weights. Some of these approaches \cite{YRZreview,Mascot22} can also describe the gapped fermionic spectrum in the anti-nodal region \cite{Shen11}. Here we shall extend theories \cite{Lee89,Kaul07,ACL08,Topo_PNAS} in which the hole pockets arise from {\it quantum\/} fluctuations of the antiferromagnetic order fractionalized into spinon excitations with spin $S=1/2$ \cite{NRSS89,NRSS90,ChubukovStarykh95,Bonetti2022}, in contrast to the SDW fluctuation theory with $S=1$ paramagnons. In this class of theories, 
the \emph{fractionalized Fermi liquid} (FL*) \cite{TSSSMV03,TSSSMV04,APAV04,Bonderson16,Assaad18,Assaad20a,ElseSenthil,Coleman22,Tsvelik24,Tsvelik25} has pocket Fermi surfaces of holes (which are holon-spinon bound states \cite{Kaul07,ACL08,Qi10,Sawatzky11,Moon11,Sawatzky11b,Mei11,Punk12,Punk15,Topo_PNAS,Grusdt18,Grusdt19,Grusdt23,Grusdt24,Balents25}), along with a background quantum spin liquid \cite{Kaul07,ACL08,Qi10,Moon11,Punk12,Punk15,Tsvelik16,ZhangSachdev_ancilla,ZhangSachdev_ancilla2,Mascot22,Joshi23,Balents25} which will be crucial to our results here. 

An important feature of the hole pockets in the FL* state is that their quasiparticles can tunnel coherently between layers, as is needed to explain the magnetotransport experiments \cite{Ramshaw20,Yamaji24}. This is in contrast to Fermi pockets of spinless holons in the holon metal state \cite{Lee89,ACL08}, for which interlayer tunneling is prohibited. Moreover,
the hole pockets were predicted to have area  $p/8$ in a FL* state \cite{Kaul07,ACL08}, yielding $1.25\%$ at $p = 0.1$—in good agreement with the area observed by the Yamaji effect. In contrast, SDW fluctuations imply area $p/4$, and consequently, we can interpret the measured Yamaji effect area \cite{Yamaji24} as a direct experimental detection of fractionalization in the cuprates \cite{Zhao_Yamaji_25}. 

We note that strong support for the above quantum interpretation of the pseudogap also comes from quantum simulations with ultracold atoms \cite{Koepsell21,Bloch24}. These show a striking difference between local multi-point correlators in the pseudogap and overdoped regimes, at temperatures without significant correlations in any broken symmetry. Moreover, the doping evolution is well-modeled by variational wavefunctions of evolution from FL* to the Fermi liquid 
\cite{Iqbal24,HenryShiwei24}.

We study a SU(2) lattice gauge theory realization of the FL* 
phase~\cite{Christos23,Christos24,Luo24,BCS24,ZhangVortex24}. While the effective action of the theory can be formulated entirely by symmetry arguments, a microscopic realization is provided by the Ancilla Layer Model (ALM) \cite{ZhangSachdev_ancilla}---see Refs.~\cite{Boulder25,ICTP25} for reviews. We employ a Monte Carlo algorithm to account for the strong thermal fluctuations 
of a SU(2) link gauge field $U$ and a charge $e$, SU(2) fundamental Higgs boson $B$ \cite{Fradkin88,LeeWen96}. When coupled to quantum electrons and 
spinons, such thermal fluctuations transform the FL* hole pockets into Fermi arcs for the photoemission spectrum. 
We note that the conversion of the hole pockets into arcs is primarily a thermal effect---in the FL* state at $T=0$, a non-zero quasiparticle residue
is present around the hole pocket even in the presence of quantum fluctuations of $B$ \cite{ZhangSachdev_ancilla,Mascot22}.

Our work is more general 
than the Born-Oppenheimer methodology used in the phase fluctuation theory, as we incorporate thermal fluctuations 
of not only the $d$-wave superconducting order parameter, but also the additional charge order---the superconducting and charge orders are interwtined as SU(2) gauge-invariant composites of the fractionalized $B$ fields.  
Furthermore, our studies suggest that quantum oscillations of the hole pockets of area $p/8$ could be observable in quantum oscillation experiments. One needs to perform experiments in clean samples at high fields and low temperatures in a regime of larger doping where the Hall coefficient remains positive and there is no field-induced charge density wave order. 

Apart from these key findings, the other important highlight of our work is demonstrating clear evidence for $h/(2e)$ 
vortices characterizing the Kosterlitz-Thouless (KT) transition to a $d$-wave superconductor. Although the $B$ boson carries 
charge $+e$, confinement leads to the formation of charge $+2e$ gauge-neutral pairs. Remarkably, each vortex core hosts 
a charge order pattern resembling STM observations by Hoffman {\it et al.} \cite{Hoffman02}.

\section{Summary of the Ancilla Layer Model}
\label{sec:ALM}

To orient the reader, and outline our paper, we begin recalling the main features of the ALM. The ALM provides an instructive and explicit derivation of the effective models studied in this paper; but, we note that it is also possible to obtain the studied models directly on phenomenological and symmetry grounds, as we discuss in Appendix~\ref{sec:fermions}.

\begin{figure}[h]
    \centering
    \includegraphics[width=3.2in]{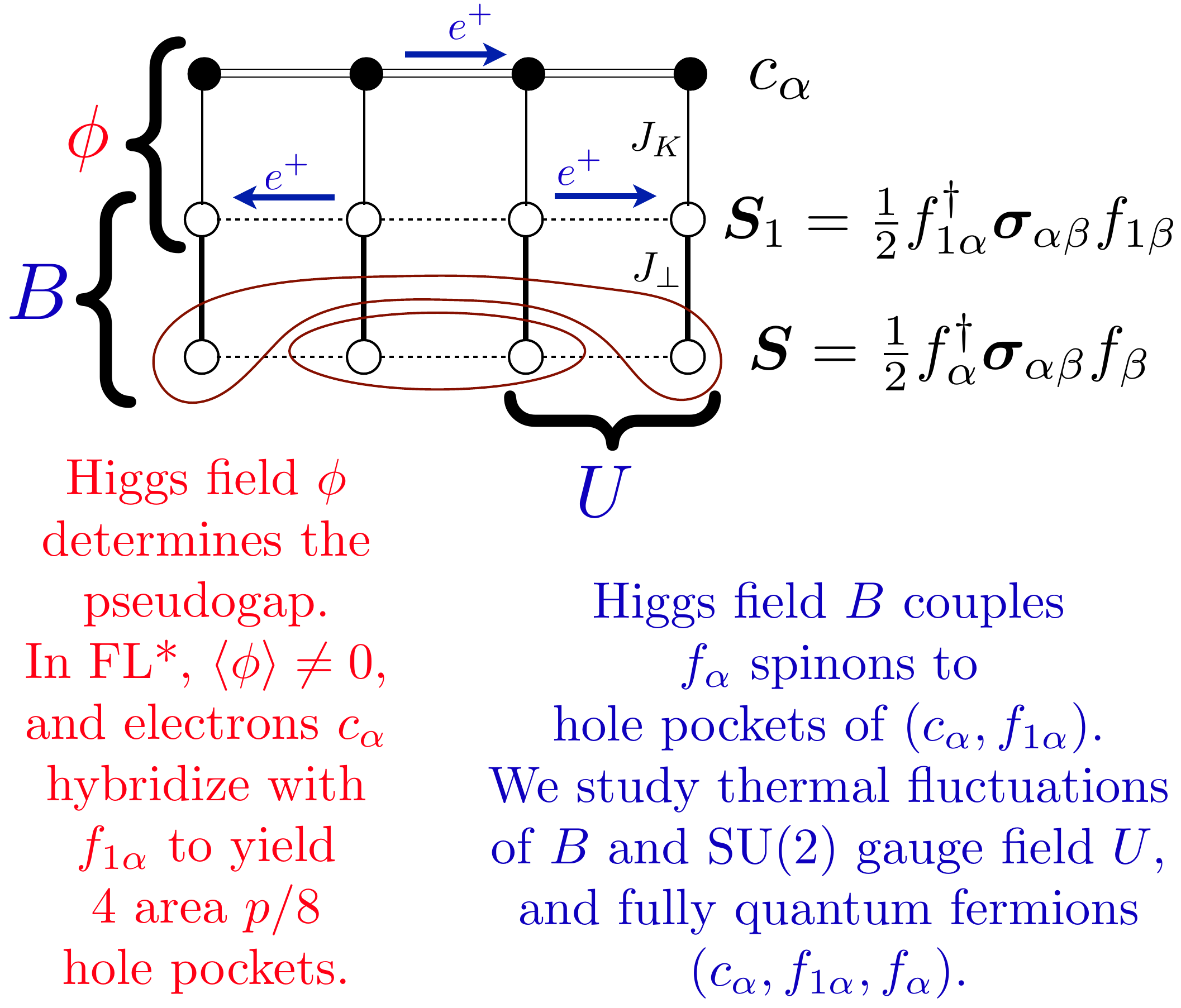}
    \caption{In the ALM, the interactions between the physical electrons $c_\alpha$ are mediated by a bilayer of $S=1/2$ ancilla spins ${\bm S}_1$ and ${\bm S}_2$ which are represented
    by fermionic spinons $f_{1 \alpha}$ and $f_\alpha$. The present paper is restricted to the underdoped regime where we fix $\phi \neq 0$ as a $c$-number.}
    \label{fig:ancilla}
\end{figure}
The key idea behind with ALM \cite{ZhangSachdev_ancilla,Boulder25,ICTP25} is to provide a mean-field theory of the pseudogap as a FL* state, while also providing a route to studying fluctuations, as we shall do here. The defining property of the FL* is the existence of electron-like quasiparticles, and so we do {\it not\/} fractionalize the electron, as is invariably done in parton theories of doped antiferromagnets. Instead, we shall fractionalize the neutral, $S=1$ paramagnon collective mode into two $S=1/2$ spins \cite{Mascot22}. This can be done systematically by coupling free electrons to a bilayer antiferromagnet with ancilla spins ${\bm S}_1$, ${\bm S}$ with rung exchange coupling $J_\perp$ (see Fig.~\ref{fig:ancilla}); then a Schrieffer-Wolff transformation at large $J_\perp$ eliminates the ancillas and yields Hubbard and longer-range interactions between the electrons \cite{Random_ancilla}.
The paramagnon is the rung triplet excitation of the antiferromagnet, which is then fractionalized into spinons $f_{1\alpha}$ and $f_\alpha$. 

Note that it is not permissible to add a single (or any odd number) layer of ancilla spins, or a single ancilla half-filled band of electrons, as is implicitly done in a number of works in the literature on the pseudogap phase. It is essential that ancilla degrees of freedom be smoothly connected to a trivial product ground state with an energy gap to maintain the correct Luttinger-type anomalies of the original Hubbard model; also, only then can the ancillas be eliminated by a Schrieffer-Wolff transformation. So any even number of $S=1/2$ spin layers are allowed, and we choose the simplest case of a bilayer.

The FL* hole pockets are obtained by hybridizing the electrons $c_\alpha$ with the $f_{1\alpha}$ with a Higgs field $\phi$, in a manner similar to the heavy Fermi liquid phase of the Kondo lattice \cite{hewson_book}; $\phi$ is obtained by decoupling the Kondo exchange interaction $J_K$. In the present paper, we will treat $\phi$ as a $c$-number constant which is determined by fitting to the anti-nodal pseudogap: see Sec.~\ref{sec:fermions} of the Supplementary Information (SI). 
Our focus here is on the influence of the bottom layer of ${\bm S}$ spins, which are assumed to form the $\pi$-flux spin liquid with a SU(2) gauge field $U$. 
Also crucial for our analysis will be a boson $B$ obtained by decoupling the $J_\perp$ interaction \cite{Christos23}: this transforms as a fundamental of the gauge SU(2) and also carries electrical charge. We summarize the quantum numbers of the dynamical fields under the gauge and global symmetries in Table~\ref{tab:quantum}.
\begin{table}[h]
    \centering
    \begin{tabular}{|c|c|c|c|}
\hline
Symmetry & Spin & Charge & Gauge SU(2)  \\
\hline 
\hline
$c_{\alpha}$ & $1/2$ & $-e$ & $0$\\
$f_{1\alpha}$ & $1/2$ & $-e$ & $0$\\
$f_{\alpha}$ & $1/2$ & 0 & $1/2$ \\
$B$ & $0$ & $+e$ & $1/2$\\
\hline
\end{tabular}
    \caption{Transformations of the dynamical fields of Fig.~\ref{fig:ancilla} under the global spin SU(2), the electromagnetic charge, and the emergent gauge SU(2).
    }
    \label{tab:quantum}
\end{table}

Our analysis will begin in Section~\ref{sec:BU} by constructing an effective energy functional for the bosons $B$ and $U$ alone. This action can be constructed solely on symmetry grounds, given the known structure of the $\pi$-flux spin liquid.
Section~\ref{sec:mean-field} will determine the mean-field phase diagram obtained by minimizing the energy functional of Section~\ref{sec:BU}.
Section~\ref{sec:mc} contains our main Monte-Carlo results of the thermal fluctuations of $B$ and $U$.

Section~\ref{sec:mc}.\ref{sec:elec} introduces the fermions: for each snapshot of $B$ and $U$, we diagonalize the Hamiltonian of the fermions. Section~\ref{sec:mc}.\ref{sec:elec} computes the zero energy electronic spectrum, while Section~\ref{sec:qo} applies a uniform magnetic field and computes the quantum oscillations in the electronic density of states.

\section{Effective energy functional for $B$ and $U$ fields}
\label{sec:BU}

The SU(2) lattice gauge theory of the pseudogap consists of 
gauge fields $U_{\vi \vj} = U_{\vj \vi}^\dagger$ residing on the links of a square lattice of sites labeled by 
$\vi \equiv (x,y)$. These are $2 \times 2$ unitary matrix obeying $U^\dagger U = \mathbf{1}$ and ${\rm det}(U) = 1$.  
The charge $e$ Higgs boson is a complex lattice doublet $B_{a\vi }$ where $a=1,2$ is the SU(2) gauge index. The energy 
functional follows entirely from the projective symmetry transformations of the underlying $\pi$-flux spin liquid 
\cite{Affleck1988,Affleck-SU2,Fradkin88} of the FL* phase \cite{Christos23,Boulder25}. These are summarized in 
Table~\ref{tab1}, along with those of the fermionic spinons. The key property
\begin{equation}
    T_x T_y = - T_y T_x \label{txty}
\end{equation}
realizes the $\pi$-flux on both the gauge-charged fermions and bosons (but $T_x$ and $T_y$ commute for all gauge-invariant observables).
\begin{table}[!htb]
    \centering
    \begin{tabular}{|c|c|c|}
\hline
Symmetry & $f_{\alpha}$ & $B_a $  \\
\hline 
\hline
$T_x$ & $(-1)^{y} f_{ \alpha}$ & $(-1)^{y} B_a $ \\
$T_y$ & $ f_{ \alpha}$ & $ B_a $ \\
$P_x$ & $(-1)^{x} f_{ \alpha}$  & $(-1)^{x} B_a $ \\
$P_y$ & $(-1)^{y} f_{ \alpha}$  & $(-1)^{y} B_a $ \\
$P_{xy}$ & $(-1)^{xy} f_{ \alpha}$ & $(-1)^{x y} B_a $ \\
$\mathcal{T}$ & $(-1)^{x+y} \varepsilon_{\alpha\beta} f_{ \beta}$  & $(-1)^{x + y} B_a $\\
\hline
\end{tabular}
    \caption{Projective transformations of the $f$ spinons and $B$ chargons on lattice sites $\vi \equiv (x,y)$ 
    under the symmetries $T_x: (x,y) \rightarrow (x + 1, y)$; $T_y: (x,y) \rightarrow (x,y + 1)$; 
    $P_x: (x,y) \rightarrow (-x, y)$; $P_y: (x,y) \rightarrow (x, -y)$; $P_{xy}: (x,y) \rightarrow (y, x)$; and 
    time-reversal $\mathcal{T}$.
    The indices $\alpha,\beta$ refer to global SU(2) spin, while the index $a=1,2$ refers to gauge SU(2). 
    }
    \label{tab1}
\end{table}

These gauge and symmetry constraints yield the needed energy functional $\mathcal{E}_2 + \mathcal{E}_4$ for the $B$ 
and $U$ fields, where
\begin{align}\label{Eq:quadraticB}
&\mathcal{E}_2 [B, U]   = \kappa \sum_{\square} 
\left[ 1- \frac{1}{2} \mbox{Re} \mbox{Tr} \prod_{\vi\vj \in \square} U_{\vi \vj} \right]
+(r+2\sqrt{2}w)\nonumber \\ 
& \times \sum_\vi B^\dagger_\vi B_\vi 
 -  i w \sum_{\langle \vi \vj\rangle} e_{\vi \vj} \left( B_\vi^\dagger U_{\vi \vj} B_\vj - B_\vj^\dagger U_{\vj \vi} B_\vi \right)~.
\end{align}
We have the standard Wilson action for the SU(2) gauge field links $U_{\vi\vj}$ whose self-coupling 
is $g$ such that $\kappa=2/g^2$ and the symbol $\square$ denotes smallest independent gauge plaquettes. 
The $B$ bosons have the familiar minimal lattice couplings to the SU(2) gauge field $U_{\vi\vj}$. This leads 
to the novel and important feature that the nearest-neighbor boson hopping is purely {\it imaginary}, $i w$. 
The $B$ bosons  experience the $\pi$-flux via the fixed field $e_{\vi\vj} =- e_{\vj \vi}$:
\begin{align}
 e_{\vi,\vi+\hat{{\bm x}}}  =  1 \,,\quad
 e_{\vi,\vi+\hat{{\bm y}}}  =  (-1)^{x}      \,, \label{su2ansatz}
\end{align}
where $\hat{\bm x} \equiv (1,0)$, $\hat{\bm y} \equiv (0,1)$. Note that the ansatz for $e_{\vi\vj}$ doubles the unit cell for fields which carry SU(2) gauge charges. But there is no doubling of the unit cell for all gauge-invariant observables, and so no breaking of translational symmetry by this ansatz.

When $U_{\vi\vj} = \mathbf{1}$, we can diagonalize $\mathcal{E}_2[B,\mathbf{1}]$ and obtain the bosonic spectrum
\begin{align}
\varepsilon_B ({\bm k}) = r + 2\sqrt{2} w \pm 2w\sqrt{\sin^2 (k_x) + \sin^2 (k_y)}\,. \label{eq:bosondisp}
\end{align}
This dispersion has minima at 2 momenta $(\pi/2, \pm \pi/2)$ in the (reduced) Brillouin zone, and the degeneracy enables 
the multiple competing orders in the Higgs phase \cite{Christos23,Boulder25}. We also need a quartic potential to stabilize $B$ fields in the condensed phase,
\begin{align}
 \mathcal{E}_4 [B, U]  & = \frac{u}{2} \sum_{\vi} \rho_{\vi}^2   + V_1 \sum_{\vi} \rho_{\vi} \left( \rho_{\vi + \hat{\bm x}} + \rho_{\vi + \hat{\bm y}} \right)  \nonumber \\
& + g \sum_{\langle \vi \vj \rangle} \left| \Delta_{\vi\vj} \right|^2 
 + J_1 \sum_{\langle \vi \vj \rangle}  Q_{\vi\vj}^2 + K_1 \sum_{\langle \vi \vj \rangle}  J_{\vi\vj}^2 \nonumber \\
&+ V_{11} \sum_{\vi} \rho_{\vi} \left( \rho_{\vi + \hat{\bm x}+ \hat{\bm y}} + \rho_{\vi +\hat{\bm x}- \hat{\bm y}} \right) \nonumber \\
& + V_{22} \sum_{\vi} \rho_{\vi} \left( \rho_{\vi + 2\hat{\bm x}+ 2\hat{\bm y}} + \rho_{\vi + 2\hat{\bm x}- 2\hat{\bm y}} \right) \,.
\label{Eq:quarticB}
\end{align}
The quartic potential has been written in terms of SU(2) gauge-invariant bilinears of $B$ with the following physical interpretations which can be deduced directly from the symmetry transformations 
in Table~\ref{tab1},
\begin{align}
&\mbox{site charge density:~}\left\langle c_{\vi \alpha}^\dagger c_{\vi \alpha}^{\vphantom\dagger} \right\rangle \sim \rho_{\vi} = B^\dagger_\vi B_\vi^{\vphantom\dagger}\,,  \nonumber \\
&\mbox{bond density:~} \left\langle c_{\vi \alpha}^\dagger c_{\vj \alpha}^{\vphantom\dagger} + c_{\vj \alpha}^\dagger c_{\vi \alpha}^{\vphantom\dagger} \right\rangle \nonumber \\ &~~~~~\sim Q_{\vi \vj} = Q_{\vj\vi} = \mbox{Im} \left(B^\dagger_\vi e_{\vi \vj}^{\vphantom\dagger} U_{\vi\vj}^{\vphantom\dagger} B_\vj \right)\,,  \nonumber \\
&\mbox{bond current:~} i\left\langle c_{\vi \alpha}^\dagger c_{\vj \alpha}^{\vphantom\dagger} - c_{\vj \alpha}^\dagger c_{\vi \alpha}^{\vphantom\dagger} \right\rangle \nonumber \\
&~~~~~\sim J_{\vi \vj} = - J_{\vj\vi} =  \mbox{Re} \left( B^\dagger_\vi e_{\vi \vj}^{\vphantom\dagger} U_{\vi \vj}^{\vphantom\dagger} B_\vj^{\vphantom\dagger} \right)\,, \label{Eq:OPs} \\
&\mbox{Pairing:~} \left\langle \varepsilon_{\alpha\beta} c_{\vi \alpha} c_{\vj \beta} \right\rangle \sim \Delta_{\vi \vj} = \Delta_{\vj \vi} = \varepsilon_{ab} B_{a\vi} e_{\vi \vj} [U_{\vi \vj}]_{bc} B_{c\vj}\,, \nonumber
\end{align}
where the symbol $\sim$ implies identical symmetry transformations of SU(2) gauge-invariant observables.
At quadratic order in $B$, all the above orderings can appear in possible, equivalent Higgs phases of $B$ fields. 
The selection between them appears at quartic order, and the couplings are chosen such that the ground state is a 
$d$-wave superconductor, and the next metastable minimum has period-4 charge density wave order, as described in Section~\ref{sec:mean-field}.

\section{Mean-Field Results}
\label{sec:mean-field}

\begin{figure}
    \centering
    \includegraphics[width=3in]{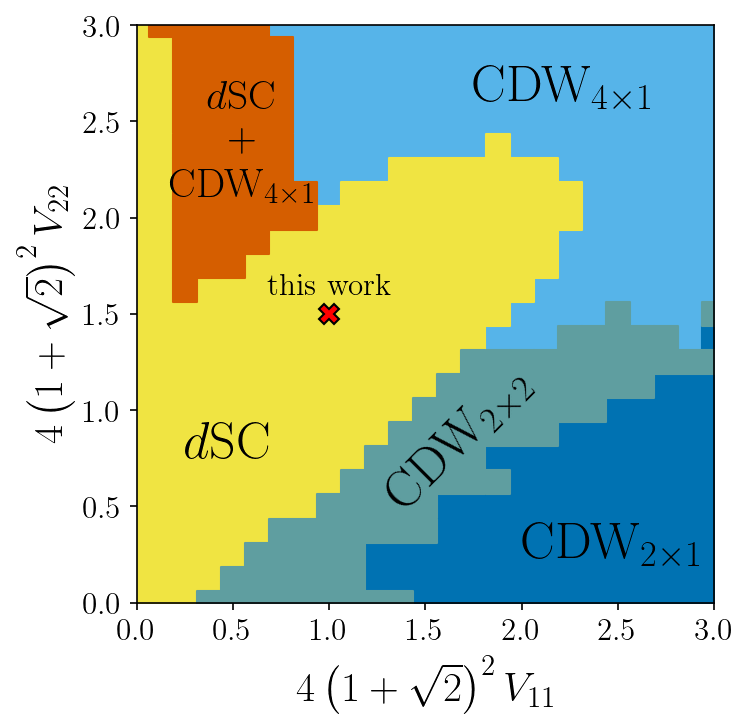}
    \caption{Mean-Field phase diagram as a function of the two couplings $V_{11}$ and $V_{22}$. The phases are labeled as follows. $d$SC stands for $d$-wave superconductivity, CDW$_{n\times m}$ stands for a charge density wave with a supercell with $n\times m$ lattice sites, $d$SC+CDW$_{4\times 1}$ is a phase with coexisting $d$-wave superconductivity and $4\times 1$ charge density wave order. The red cross marks the parameter values chosen for the Monte Carlo simulations discussed in the main article.}
    \label{fig:MF phase diagram}
\end{figure}

We start by analyzing the energy functionals in Eqs.~\eqref{Eq:quadraticB} and \eqref{Eq:quarticB} within a mean-field scheme. There is no constraint equating the magnitude of $B$ to $p$, unlike earlier works \cite{Fradkin88,LeeWen96}. We work in the limit $\kappa\to \infty$, where gauge field fluctuations are strongly penalized, which allows us to take $U_{\vi\vj}=\mathbbm{1}$. We then minimize the functional $\mathcal{E}_2[B,U=\mathbbm{1}]+\mathcal{E}_4[B,U=\mathbbm{1}]$ with respect to $B_\vi$ on a finite lattice with size $N\times N$ and periodic boundary conditions. Note that, due to the short-range nature of the energy functional and to the mean-field approximation, increasing $N$ has the only effect of allowing for longer periods of the spatial modulations of $B_\vi$. For the model parameters we will be discussing in the following, we found that $N=8$ was sufficient to obtain the lowest energy state. We fix the parameters in the functional to
\begin{eqnarray}
    && r=-0.732, ~w=0.40, ~u=0, ~V_1=0, ~g=0.021446, \nonumber \\
    && J_1=~K_1=\frac{2}{4(1+\sqrt{2})^2}\,,
    \label{eq:pars1}
\end{eqnarray}
and calculate the mean-field phase diagram as a function of $V_{11}$ and $V_{22}$. We classify the different phases according to the order parameters listed in Eqs.~\eqref{Eq:OPs}. Although not present for the chosen parameter values, we expect that pair density wave (PDW) states are also possible in our mean-field theory \cite{Christos23}.

\begin{figure}
    \centering
    \includegraphics[width=3in]{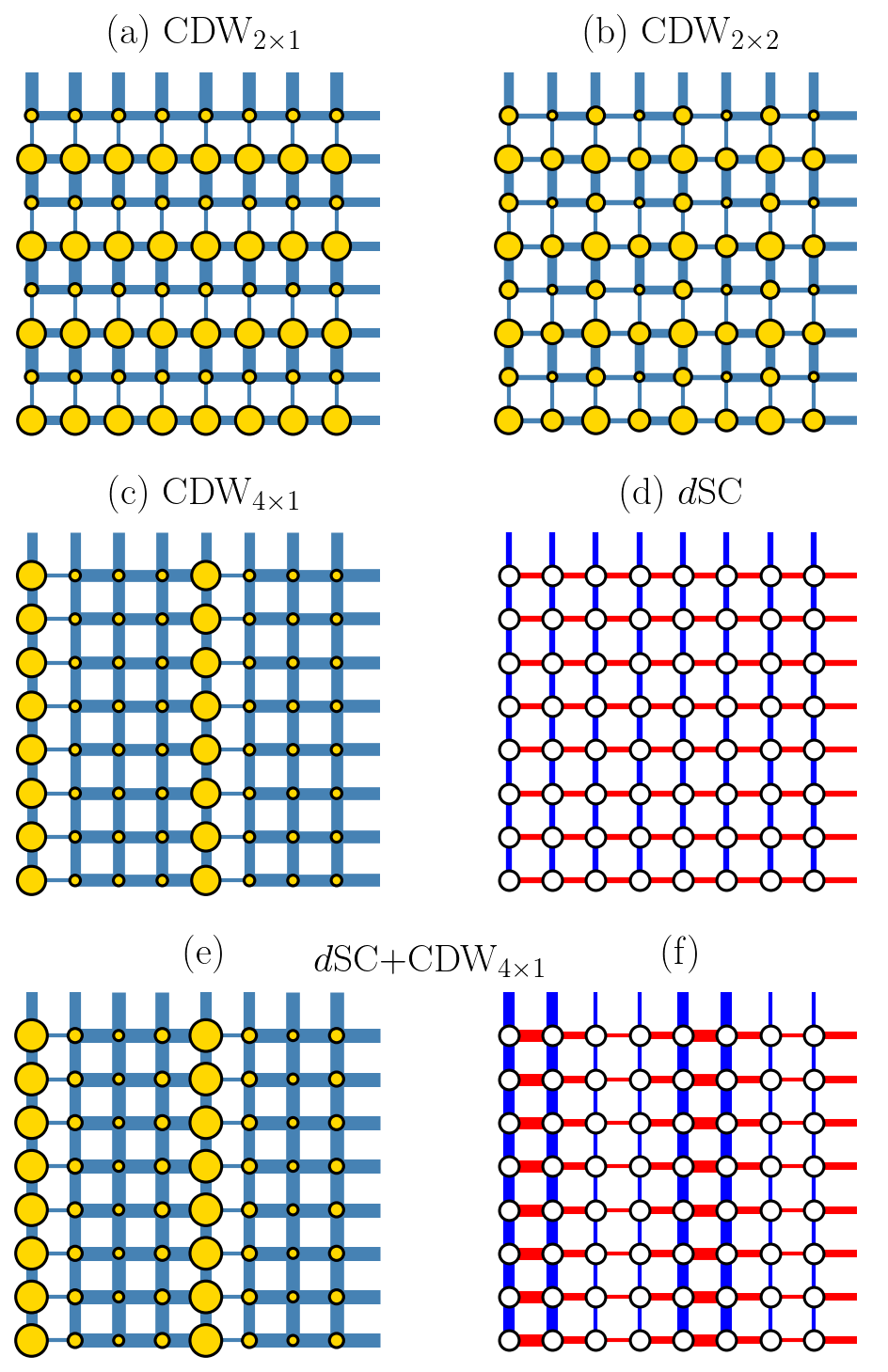}
    \caption{Charge, bond and pairing density profiles of the phases found in Fig.~\ref{fig:MF phase diagram}. (a-c): Charge and bond densities of $2\times1$, $2\times2$, $4\times 1$ CDW phases, respectively. Larger (smaller) bullets indicate a higher (lower) onsite charge density, whereas thick (thin) lines indicate a higher (lower) bond density. (d) Bond pairing density of the $d$-wave superconducting state. Red (blue) lines represent positive (negative) bond pairing amplitudes. Here, all lines have the same thickness as pairing is uniform. Charge and bond densities (e) and pairing bond density (f) for the state where $d$-wave superconductivity coexists with a $4\times 1$ charge density wave. The loop currents $J_{\vi\vj}$ are zero for all phases.}
    \label{fig:phases.}
\end{figure}

In Fig.~\ref{fig:MF phase diagram}, we show the mean-field phase diagram as a function of the couplings $V_{11}$ and $V_{22}$. We find a phase hosting uniform $d$-wave superconductivity, in which $\Delta_{\vi\vj}$ is nonvanishing and obeys the relation $\Delta_{\vi,\vi+\bm{x}}=-\Delta_{\vi,\vi+\bm{y}}=\Delta_0$ for all sites $\vi$. We then find three distinct charge density wave phases in which the site and bond densities are spatially modulated in space with supercells of sizes $2\times 1$, $2\times 2$, and $4\times 1$. Finally, we find a state in which $d$-wave superconductivity coexists with a $4\times 1$ CDW. Fig.~\ref{fig:phases.} shows the charge, bond and pairing bond densities of all of the above-mentioned phases.

In the Monte Carlo simulations, we used the parameter set in Eq.~(\ref{eq:pars1}), and as indicated in Fig.~\ref{fig:MF phase diagram}, we employ the magnitude of next-to-nearest-neighbor interactions 
\begin{eqnarray}
   V_{11}= \frac{1}{4(1+\sqrt{2})^2}=0.04289, ~V_{22}=0.0643 \,, \label{eq:pars2}
\end{eqnarray}
where the ground state is a $d$-wave superconductor. The next higher energy state is a $4\times 1$ CDW, and this will influence the structure of the vortex core.


\section{Monte Carlo Results}
\label{sec:mc}

Our interest in this paper is limited to the intermediate temperature pseudogap regime and its transition to superconductivity upon lowering temperature. As long as we remain far from any quantum phase transition at lower temperatures, this allows us to limit consideration to only thermal fluctuations of the bosonic fields $B$ and $U$. The fermionic excitations, which will be considered in Section~\ref{sec:elec}, will however be treated quantum mechanically exactly. We are therefore following a Born-Oppenheimer procedure, with $B$ and $U$ playing the role of nuclear positions in molecules, similar to that followed in the phase fluctuation approach \cite{Franz98,Scalapino02,Dagotto05,Berg07,Li_2010,Li11JPhys,Li11PRB,Li_2011,Sumilan17,Majumdar22,YQi23,Xiang24,yang2025}. For this, we only need the energy functional $\mathcal{E}_2 [B,U] + \mathcal{E}_4 [B,U]$ and do not need to introduce time derivative terms in the bosonic action \cite{Christos23} to account for quantum fluctuations. Some computations which include time derivatives and quantum fluctuations $B$ at the one-loop level are presented in Appendix~\ref{app:oneloop}, and the results are similar to the Monte Carlo results below.

We performed a Monte Carlo calculation of the classical partition function
\begin{align}
& \mathcal{Z}_{2+0} = \int \prod_\vi \mathcal{D} B_\vi \int \prod_{\langle \vi\vj \rangle} \mathcal{D} U_{\vi\vj} \nonumber \\
&~~~~~~~~~~~~~~~~~~~~\times\exp \left[- (\mathcal{E}_2 [B, U] + \mathcal{E}_4 [B, U])/T \right] 
\label{eq:Z20}
\end{align}
on a two-dimensional $N \times N$ lattice where $N=64,96,128,192$. The inverse gauge coupling $\kappa=1$ is set such 
that we are in a weakly coupled regime, not very far from the mean-field, yet with significant gauge fluctuations. The 
initial conditions were chosen such that $B_\vi$ are random numbers on each lattice site and the gauge links $U_{\vi\vj}$ 
are $2\times 2$ identity matrices on each bond connecting the sites $\vi$ and $\vj$. Our algorithm involves 
optimizing the energy at each temperature $T$  with a Metropolis accept/reject criterion. This is 
performed by updating the chargon field $B_\vi$ on each lattice site and gauge links $U_{\vi\vj}$ on 
all the bonds during one sweep. We measure the value of $\mathcal{E}_2[B,U]+\mathcal{E}_4[B,U]$ after each update 
and then accept if the difference in energy $\Delta \mathcal{E}$ compared to the previous step is either negative 
or if it is positive but $\mathrm{exp}(-\Delta \mathcal{E}/T)$ is larger than a random number chosen in the interval 
$[0,1]$, and reject otherwise. From the plot of the energy as a function of the number of sweeps, we monitor the onset of 
a plateau where the energy stabilizes, which ensures that our algorithm has achieved thermalization. Next, we consider 
thermalized configurations that are sufficiently decorrelated, separated by $3$-$4$ times the typical 
autocorrelation time and perform thermal averages of the various order parameters at each value of $T$. 
Further results on the fluctuations of the superconducting and charge density wave orders across $T_c$ appear 
in Appendix~\ref{sec:KT}.

\begin{figure*}
    \centering
        \includegraphics[width=0.75\textwidth]{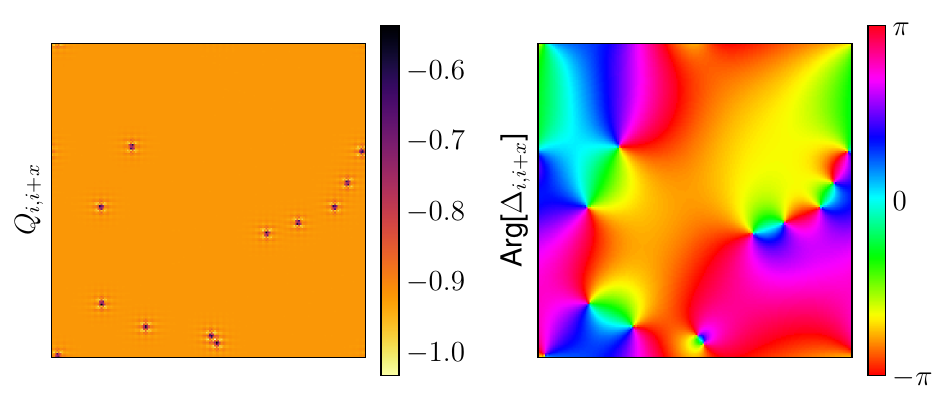}
    \caption{The bond density $Q_{\vi,\vi+\hat{\bm x}}$ (left panel) and the distribution of the phase of the superconducting order 
    parameter $\Delta_{\vi,\vi+\hat{\bm x}}$ (right panel) on a $192\times 192$ real-space lattice for the ground state at temperature $T=10^{-8}$.}
    \label{fig:BondDensPairingPhLowT}
\end{figure*}

\begin{figure}
    \centering
\includegraphics[width=0.35\textwidth]{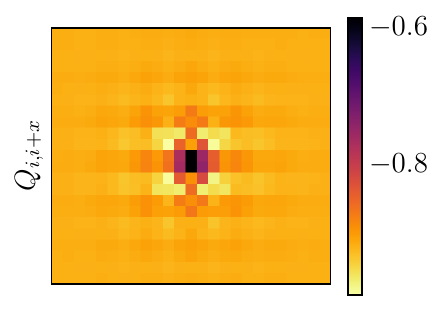}
\caption{A zoomed-in view of the bond density order parameter $Q_{\vi,\vi+\hat{\bm x}}$ in Eq.~(\ref{Eq:OPs}) in the real-space vicinity of a vortex core.}
    \label{fig:ZoomedInVortex}
\end{figure}

In Fig.~\ref{fig:BondDensPairingPhLowT} we show a snapshot of the low temperature superconducting state of this theory. We particularly draw attention towards the vortices shown in the right panel: the phase of the SU(2) gauge-invariant superconducting order parameter $\Delta_{\vi \vj}$ defined in Eq.~(\ref{Eq:OPs}), winds by $ 2\pi$. Since $\Delta_{\vi \vj}$ is bilinear in the $B_\vi$ which carries electrical charge $2e$, these vortices will carry a flux $h/(2e)$ upon including the electromagnetic gauge fields. It is notable that such vortices appear even though the matter fields $B$  carry charge $e$, this is a direct consequence of the confinement of the SU(2) gauge fields~\cite{ZhangVortex24}. The 
left panel of Fig.~\ref{fig:BondDensPairingPhLowT} shows a snapshot of the same vortices with a charge order around the core of each vortex. A zoomed-in view of a typical vortex is shown in Fig.~\ref{fig:ZoomedInVortex}.
A period-4 checkerboard modulation, far from the vortex core arising from the choice in Eq.~(\ref{eq:pars2}), 
is evidently distorted near the center of the vortex. The vortex core induces an SU(2) gauge flux \cite{SSvortex}, 
and re-orients $B_\vi$ fields to induce a period-4 charge order,  the mechanism of which was discussed in 
Ref.~\cite{ZhangVortex24}. 

\subsection{Superconducting phase transition}
\label{sec:sc}

\begin{figure}
    \centering
        \includegraphics[width=0.44\textwidth]{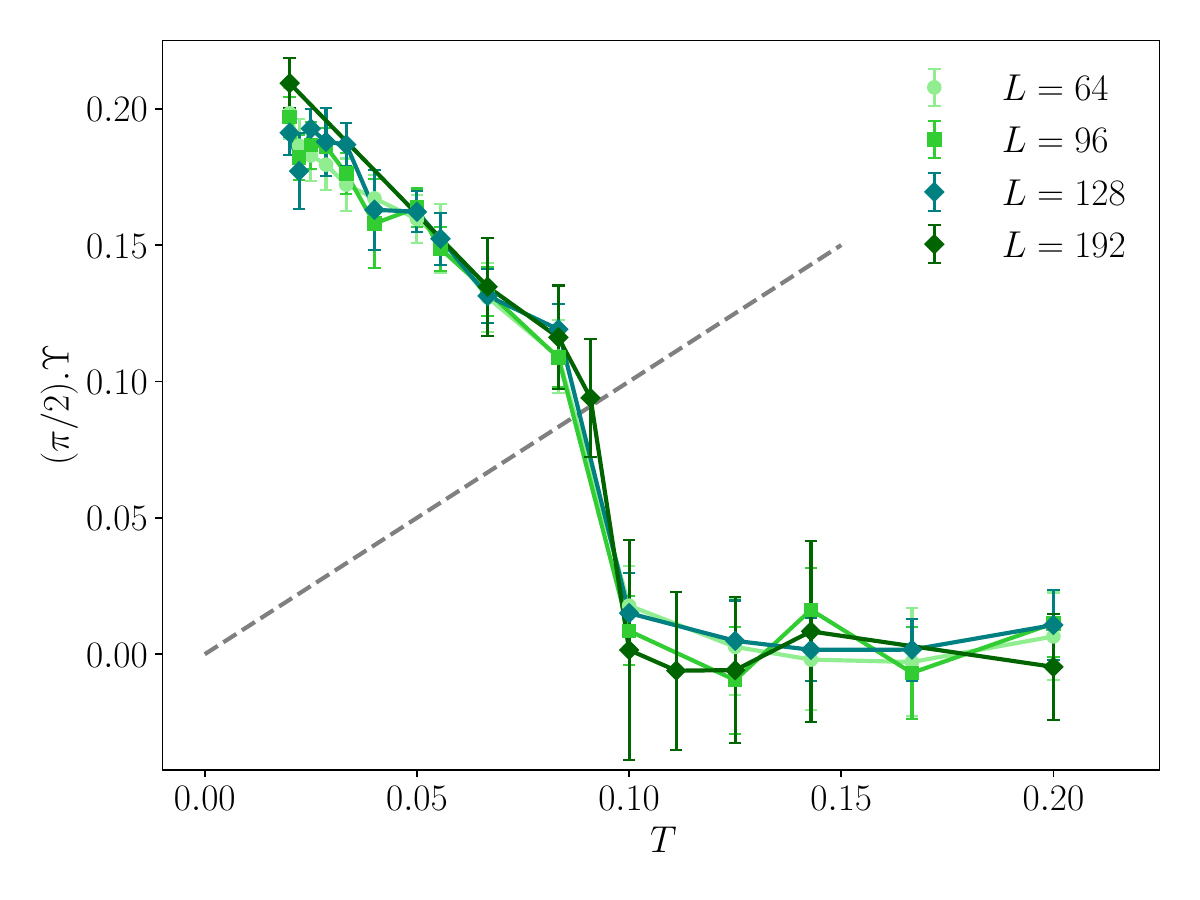}
           \caption{The data points for the average helicity modulus $\Upsilon$ as a function of temperature for different lattice volumes, connected by lines for visual clarity. The dashed line has a slope of unity and its point of intersection with the data curves gives the temperature corresponding to the Kosterlitz-Thouless transition in Eq.~(\ref{Eq:TKT}).}
    \label{fig:helicityMod}
\end{figure}

Upon increasing the temperature, we observe a proliferation of the vortices in Fig.~\ref{fig:BondDensPairingPhLowT}. We expect a Kosterlitz-Thouless (KT) transition to the normal state, similar to that in an XY model for the phase of the superconducting order parameter.  To test for such a transition, we measure the helicity modulus $\Upsilon$. This is defined by introducing a \emph{fictitious} U(1) gauge field $A_{\vi\vj} = - A_{\vj\vi}$ which acts on the electromagnetic charge of the $B_\vi$, modifying the hopping term in Eq.~(\ref{Eq:quadraticB}) to 
\begin{align}
   i w \sum_{\langle \vi \vj\rangle} e_{\vi \vj} \left( B_\vi^\dagger U_{\vi \vj} e^{-i A_{\vi\vj}} B_\vj - B_\vj^\dagger U_{\vj \vi} e^{-i A_{\vj\vi}} B_i \right)\,. \label{BAhop}
\end{align}
Note that U(1) gauge invariance also requires coupling the quartic terms proportional to \( J_1 \) and \( K_1 \) in Eq.~\eqref{Eq:quarticB} to the U(1) gauge field. However, one can show that this coupling vanishes when \( J_1 = K_1 \), as is the case in our numerical calculations.
We chose
\begin{align}
    A_{\vi, \vi+\hat{{\bm x}}} = \Theta, ~ A_{\vi, \vi+\hat{{\bm y}}} = 0 
\end{align}
and this induces a twist of $L \Theta$ in the boundary conditions for $B_\vi$ around the $x$-loop of the spatial torus. With this twist, one can obtain the helicity modulus from the following relation,
\begin{eqnarray}
   4 \Upsilon = \lim_{\Theta \rightarrow 0} \frac{2  \Delta F}{L^2 \Theta^2}~.
   \label{eqn:HelicityModDef}
\end{eqnarray}
The quantity $\Delta F$ represents the difference in the free energy at each value of the inverse temperature 
$\beta=1/T$,
\begin{displaymath}
 \Delta F = -\frac{1}{\beta} ~\ln ~\Big \langle \mathrm{exp} \left[-\beta~(\mathcal{E}(\Theta) - \mathcal{E}(\Theta=0)) \right] \Big \rangle_{\mathcal{H}(\Theta=0)}~,
\end{displaymath}
where $\mathcal{E} = \mathcal{E}_2 + \mathcal{E}_4$. The leading factor 4 in Eq.~(\ref{eqn:HelicityModDef}) is crucial; 
it arises due to the fact that the SU(2) gauge invariant order parameter $\Delta_{\vi,\vj}$ carries charge $2e$, in 
contrast to the charge $e$ carried by $B_\vi$ in Eq.~(\ref{BAhop}). While calculating $\Delta F$ the ensemble averaging 
was performed with respect to the gauge field configurations without twist. The values of the helicity modulus as a function of temperature for different lattice sizes are summarized in Fig.~\ref{fig:helicityMod}. The helicity modulus shows a jump around the temperature associated with Eq.~(\ref{Eq:TKT}), supporting the existence of a KT transition.  In our simulations, the XY order parameter is a composite of the underlying degrees of freedom, and the fluctuations of the $U_{\vi\vj}$ needed to be carefully equilibrated to realize a KT transition of a 
charge $+2e$ order parameter, especially for larger system sizes. The transition temperature $T_{KT}$ can be determined from 
the Nelson-Kosterlitz criterion \cite{PhysRevLett.81.5418}
\begin{equation}
    \frac{\pi}{2}~\Upsilon(T=T_{KT})=~T_{KT}\,. 
    \label{Eq:TKT}
\end{equation}
which is close to $T_{KT}\simeq 0.09 $. Our observation of $ \pm 2 \pi$ vortices in the charge $2e$ parameter, and 
the rapid variation of $\Upsilon$ in Fig.~\ref{Eq:TKT} around the dashed line thus unambiguously support the presence 
of a KT transition in this system.

\subsection{Electronic spectral weight at zero energy}
\label{sec:elec}

\begin{figure*}
    \centering
    \includegraphics[width=0.8\linewidth]{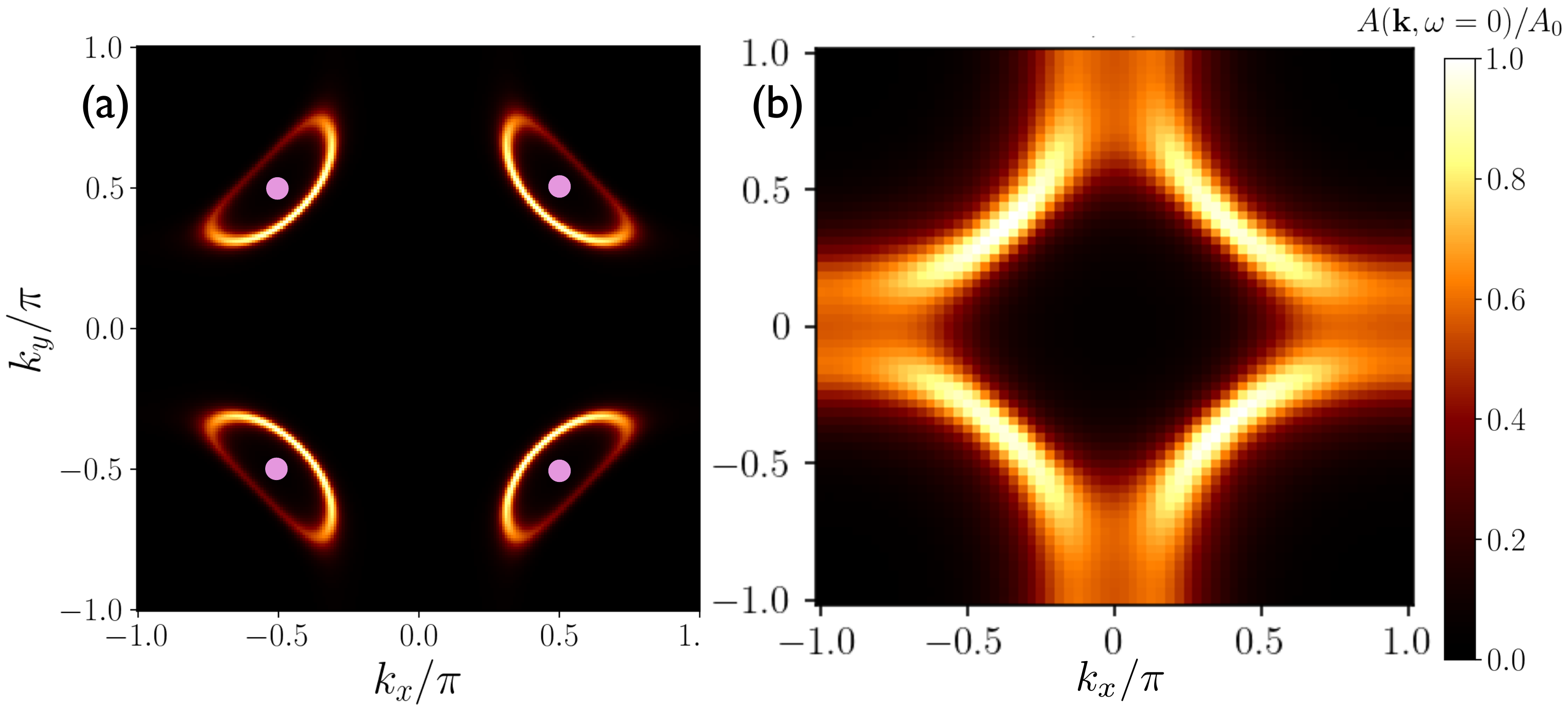}
    \caption{Zero frequency electron spectral weight (a) without and (b) with the coupling to spinons via thermal fluctuations of $B,U$ for $T>T_c$. The nodal spinons are the pink circles in (a), and these couple efficiently to the backsides of the pockets. The hole pocket of area $p/8$ in (a) transforms to a Fermi arc in (b). Parameters are specified in Appendix~\ref{sec:fermions}.}
    \label{fig:Photo_before_after}
\end{figure*}
We now couple the thermal ensembles of $B$ and $U$ fields to the fermionic spinons and 
electrons in order to study the effects of thermal fluctuations on the electronic spectrum as measured by photoemission. The fermionic Hamiltonian 
is dictated by gauge invariance and the transformations in Table~\ref{tab1}: its derivation is reviewed elsewhere \cite{Boulder25,ICTP25}, and the full form in the Ancilla Layer Model (ALM) is presented in Appendix~\ref{sec:fermions}. 
We choose the Yukawa couplings between $B$ and fermions in Eq.~(\ref{Yukawa}) as $g_1=0$ and $g_2=1$, so that the only direct coupling is between the bottom two layers of the ALM; these two layers have a large rung exchange interaction, $J_\perp$, between them in the ALM. All other parameters are kept similar to those described in Ref.~\cite{Mascot22} with a doping $p=0.2$.

For each thermal realization of $B$ and $U$ fields, we diagonalize the fermion Hamiltonian and compute the thermally averaged Green's functions. The results for the spectral weight at $\omega=0$ in the momentum space are compared with the corresponding quantity at the mean-field level in 
Fig.~\ref{fig:Photo_before_after}. 
\begin{figure}
    \centering
    \includegraphics[width=0.95\linewidth]{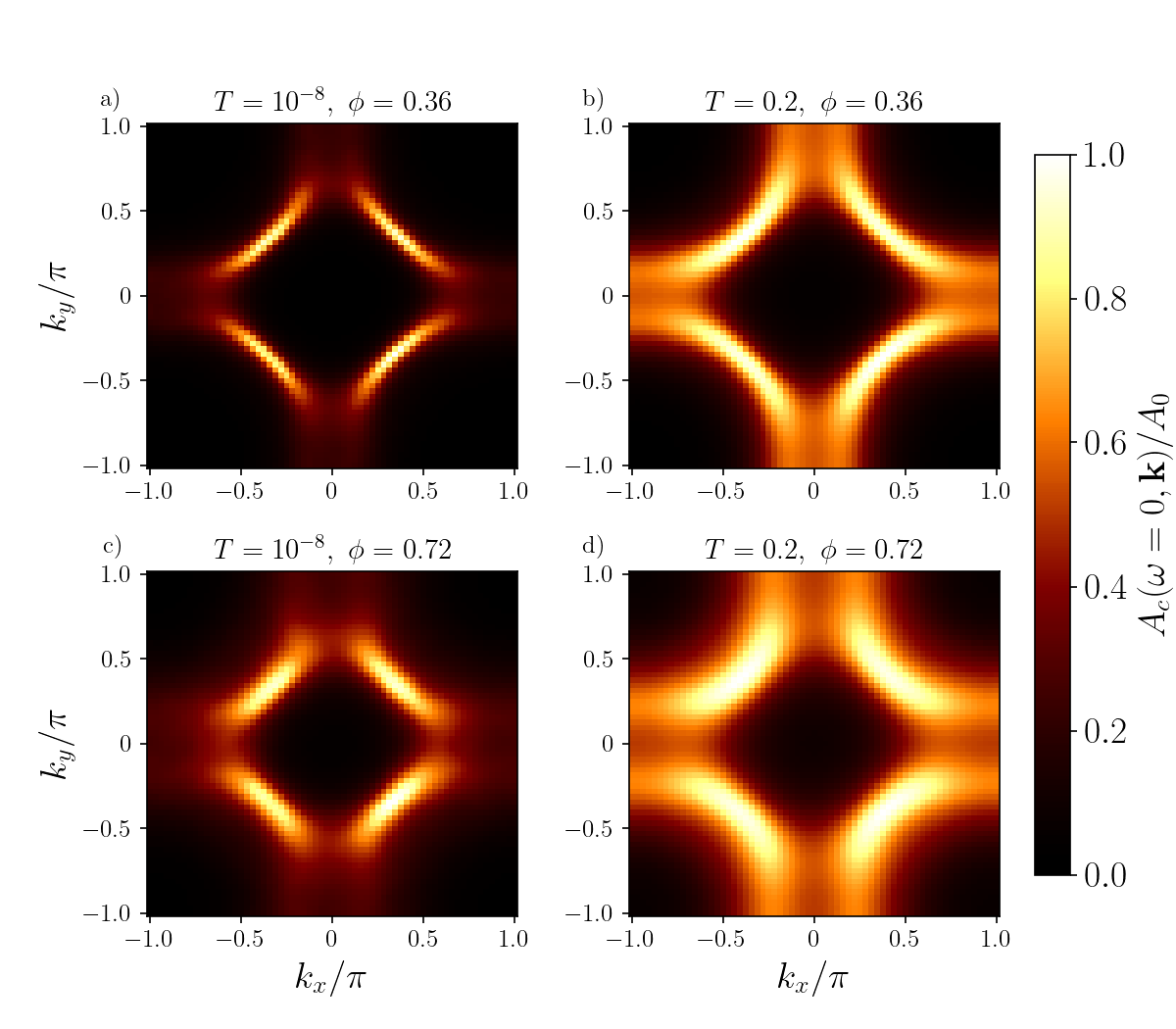}
    \caption{Zero frequency electron spectral weight of $\mathcal{H}[f] + \mathcal{H}[c,f_1] + \mathcal{H}[c,f,f_1]$ averaged over the thermal Monte-Carlo ensembles of $B$ and $U$ fields. The parameters are given in Eqs.~(\ref{eq:pars1}) and (\ref{eq:pars2}). The temperatures are set to $T=10^{-8} (\ll T_c)$ for panels (a,c) and $T= 0.2~(>T_c)$ for panels (b,d). The hybridization $\phi$ between the top two layers of fermions, which determines the magnitude of the pseudogap near $(\pi,0)$, $(0,\pi)$, has values $\phi = 0.36$ for panels (a,b) and $\phi=0.72$ for panels (c,d). We choose spin liquid hopping and broadening parameter to be $J=0.2/\sqrt{2}$ and $\eta=0.01$ respectively. All other parameters in the fermion Hamiltonian defined in Eqs.~(\ref{Top2layers_Hamiltonian}) are kept similar to ones described in Ref.~\cite{Mascot22}. The figure in (b) is the same as that in Fig.~\ref{fig:Photo_before_after}b.}
    \label{fig:fermionMC}
\end{figure}

Remarkably, the pocket backsides are no longer visible in the presence of $B,U$ fluctuation-mediated coupling to spinons, and the spectra are similar to the observed Fermi arcs \cite{Norman98,ShenShen05,Johnson11,Kondo20,Kondo23,Damascelli25}.
This arises from an effect similar to that found in the $d$-wave superconducting ground state in Ref.~\cite{Christos24}: the Yukawa coupling $g_2$ between $f_1$, $f_2$, $B$ in Eq.~(\ref{Yukawa}) hybridizes the electronic quasiparticles on the pocket backsides (which are dominantly $f_1$) with the spinons $f_2$. In the presence of a thermally fluctuating $B$ field, this is sufficient to remove the backside spectral intensity. 

We also examine in Fig.~\ref{fig:fermionMC} the effect of varying the parameter $\phi$ in Eq.~(\ref{Top2layers_Hamiltonian}), which determines the pseudogap in the anti-nodal region. As expected \cite{Christos24}, increasing $\phi$ enhances the nodal character of the spectrum at lower values of $T$. We observe the shrinking of the Fermi arcs below $T_c$ in Fig.~\ref{fig:fermionMC}, and they will eventually become the nodal quasiparticles of Ref.~\cite{Christos24} at $T=0$.

\section{Quantum Oscillations}
\label{sec:qo}

\begin{figure*}
    \centering
    \includegraphics[width=6.5in]{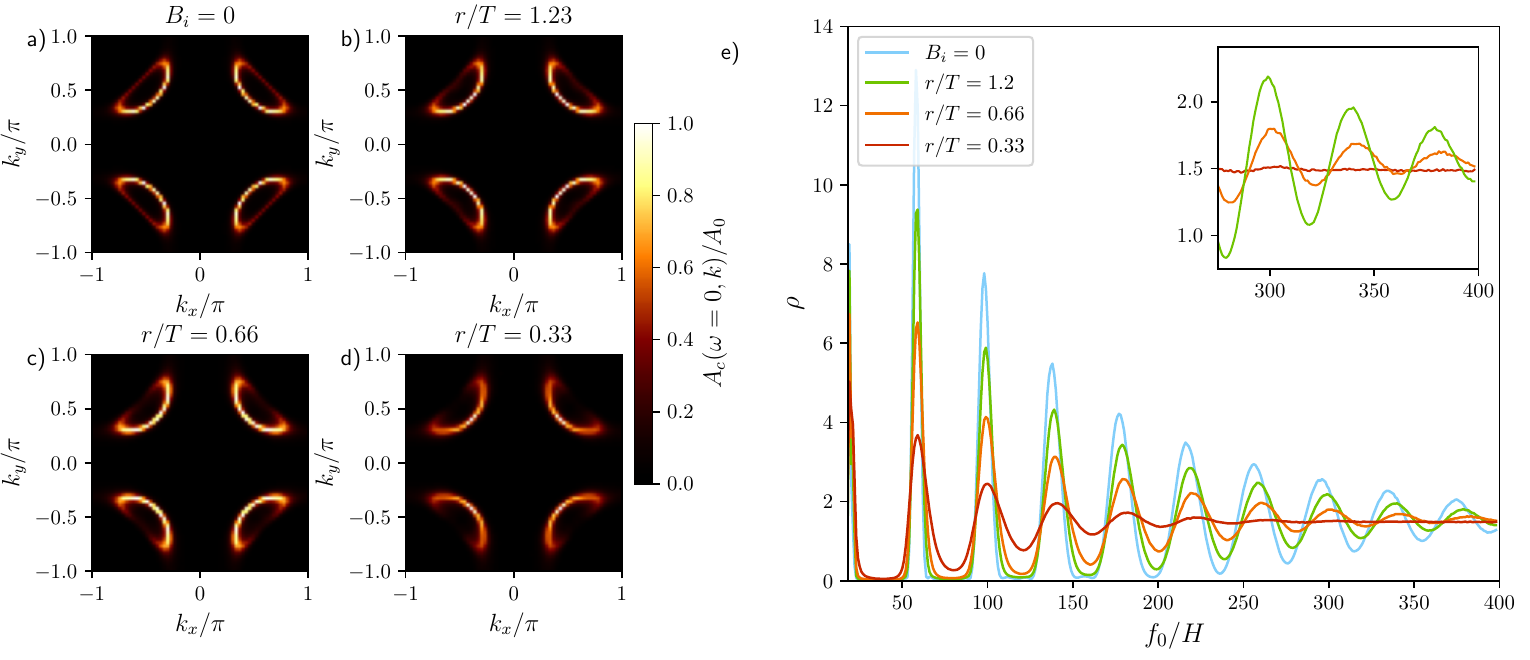}
    \caption{(a-d) Zero frequency electronic spectral weight in the presence of gaussian distributed thermal fluctuations  of $B$ mediating a coupling to spinons. 
    We use only the gaussian contributions to the quadratic free energy Eq.~(\ref{Eq:quadraticB}) about the saddle point $U_{\vi\vj}=1$ with $w=0.2/(2\sqrt{2})$ and averaged over 100 samples each with a broadening parameter $\eta=0.01$. The spin liquid hopping is $J=0.2/\sqrt{2}$. All other model parameters used are the same as Fig.~\ref{fig:fermionMC}. 
    (e) The density of states $\rho$ as a function of the inverse of the magnetic field $1/H$ for the parameter values of the plots in (a-d). The inset is an expanded view of the data at smaller values of $H$.  The frequency $f_{0}={h}/({ea_0^2})$ corresponds to the area of the Brillouin zone; for the cuprates $f_0 \approx 28600$ T.}
    \label{fig:Photo_Osc}
\end{figure*}

We next compute the quantum oscillations in the electronic density of states in presence of 
thermal fluctuations and an external magnetic field $H$. Unlike photoemission, this observable does not involve the ejection of electrons from the sample, and so its behavior can be distinct from the Fermi arc spectrum computed above.
As large system sizes are required to be sensitive to the small 
Fermi pockets of size $p/8$, we have implemented only those fluctuations of $B$ fields which are of gaussian type in this 
computation; we also set $U=1$ in this computation, equivalent to taking the limit $\kappa \rightarrow \infty$ in Eq.~(\ref{Eq:quadraticB}). For details, see Appendix~\ref{sec:Gaussianfermion}. Our findings are summarized in 
Fig.~\ref{fig:Photo_Osc}. 

These results in Fig.~\ref{fig:Photo_Osc}
demonstrate that the density of states can exhibit quantum oscillations of a small Fermi surface with an area fraction of the total Brillouin zone equal to ${p}/{8}$, even when the electronic spectral intensity of the back side of the pockets is diminished by thermal fluctuations.
Consequently, our results indicate that other experimental probes which characterize the area enclosed by the Fermi surface \cite{Zhao_Yamaji_25,FuChun25,Sondheimer26}, including the recent experiments which probed the Yamaji effect \cite{Yamaji24}, should also detect an area $p/8$.

Kunisada {\it et al.\/} \cite{Kondo20} have observed quantum oscillations of hole pockets in the lightly-doped regime with SDW order, where the area of the hole pocket is $p/4$. The FL* state has area $p/8$, but the values of $p$ are larger than in the SDW state: so the field range needed should be the similar to those of Kunisada {\it et al.\/}. A caveat is that these field ranges correspond to the largest values of $f_0/H$ shown in the inset of Fig.~\ref{fig:Photo_Osc}b, and the oscillations are significantly suppressed at these fields for parameter values for which the arcs are well-formed in Fig.~\ref{fig:Photo_Osc}a. But we note that gauge fluctuations of $U$ (not included in Fig.~\ref{fig:Photo_Osc}) further enhance the arc formation.

Turning to thermodynamic observables, we do not present results including gaussian fluctuations of $B$. But, for reference, we consider the case of free fermions, for which quantum oscillations in transport or thermodynamics have an additional Lifshitz-Kosevich factor \cite{Shoenberg1984}, which in turn depends on the effective mass of the quasiparticles. 
Kunisada {\it et al.\/} \cite{Kondo20} determined a hole quasiparticle mass from quantum oscillations of $m^* \approx 0.7 m_e$, where $m_e$ is the bare electron mass, in a regime with SDW order. We expect a similar hole mass in FL* \cite{Joshi23}. We therefore obtain the thermal damping factor \cite{Shoenberg1984}
\begin{equation}
R_T = \frac{X}{\sinh X} \,, \quad X = \frac{2 \pi^2 k_B T m^*}{\hbar e B} \approx \frac{10.3 T}{B}\,,
\end{equation}
where $T$ is measured in Kelvin, and $B$ in Tesla. The rather light hole mass observed in Ref.~\cite{Kondo20} makes for relatively favorable conditions to find a sample with $T$ and $p$ large enough and $B$ small enough to avoid charge order, but $T$ small enough and $B$ large enough to have an appreciable $R_T$. 

The small value of $m^\ast$ is also supported by the Chan {\it et al.} observation of the Yamaji effect \cite{Yamaji24}, which requires a Fermi energy larger than the observation temperature of 80 K. Estimating $k_F$ from the Fermi surface area, $\pi k_F^2 = (p/8) (2 \pi/a)^2$ (where $a$ is the square lattice spacing), and using $m^\ast$ from  Kunisada {\it et al.\/}, we obtain a Fermi energy $\hbar^2 k_F^2 /(2 m^\ast) = 690$~K. Chan {\it et al.} also estimate $\omega_c \tau = 2.6$, where $\tau$ is a transport time. Quantum oscillations are sensitive to the shorter quasiparticle scattering times, and so clean samples will be necessary to avoid suppression by the Dingle factor.


\section{Perspective}

The FL* theory posits a quantum origin of the fermionic pseudogap in the anti-nodal region of the Brillouin zone, not arising due to the fluctuations of any underlying order parameter. Characteristic features of FL* are the presence of hole pockets of fractional area $p/8$ and a \emph{background} spin liquid; Christos {\it et al.\/} \cite{Christos23} argued that the appropriate spin liquid was one with massless fermion Dirac spinons. Here we have shown that thermal fluctuations of this spin liquid couple the spinons and electrons in a manner which converts the photoemission and STM spectrum to Fermi arcs, see Fig.~\ref{fig:Photo_before_after}. We emphasize the thermal nature of this conversion, as the FL* state at $T=0$ has a non-zero quasiparticle residue around the hole pocket, even after including the coupling to spinons \cite{Mascot22}.

We have also demonstrated that the pocket-like nature of the quasiparticle excitations of FL* can show up in quantum oscillations in the presence of gaussian thermal fluctuations, see Fig.~\ref{fig:Photo_Osc}. This justifies the pocket interpretation of recent magnetotransport experiments \cite{Ramshaw20,Yamaji24}. Our studies indicate that 
it may be possible to observe quantum oscillations of the hole pockets of fractional area $p/8$ in sufficiently high fields, low temperatures, and clean samples which do not 
have field-induced charge order. Existing measurements of quantum oscillations in the lightly-hole-doped cuprates are in samples with charge density wave order which leads to the formation of electron pockets \cite{Sebastian_review,Zhang_Mei_2016,BCS24}. The presence of electron pockets is signaled by a negative Hall co-efficient, whereas the oscillations from the hole pockets considered here should have a positive Hall co-efficient.

A specific microscopic theory of FL* in a single-band model is provided by the Ancilla Layer Model (ALM) \cite{ZhangSachdev_ancilla}, which we specified in Appendix~\ref{sec:fermions}. Such an ALM has been used to formulate variational wavefunctions which have compare well to local multi-point correlators measured in ultracold atom quantum simulators  \cite{Iqbal24,HenryShiwei24,Koepsell21}. 

Our results also provide a new perspective on the role of intertwined orders in the pseudogap. In contrast to Landau theory approaches which work directly with a free energy for the orders \cite{KivelsonRMP,Fradkin25}, we have a fractionalized Higgs field $B$ whose gauge-invariant composites describe the intertwined orders. Although our approach maps onto the Landau theory after integrating out the thermal SU(2) gauge field, the fractionalized gauge theory has the advantage of allowing a description of the fermionic spectrum in terms of states of a Hamiltonian with thermally fluctuating orders in a local manner. 
Moreover, as a consequence partly due to
Eq.~(\ref{txty}), there is no possible orientation of $B$ without a broken symmetry, and hence the underdoped vortex core cannot be that found in overdoped region \cite{Renner21}---the latter is compatible with conventional Bardeen-Cooper-Schreiffer theory \cite{WM95}. We showed that there is a 
natural choice of parameters for which the underdoped vortex charge order halos discovered by Hoffman {\it et al.\/} \cite{Hoffman02} appear 
in our theory while including full thermal gauge fluctuations.

\subsection*{Acknowledgements}

We thank Mun Chan, Andrey Chubukov, Antoine Georges, Neil Harrison, Steven Kivelson, Gabriel Kotliar, Bertrand Halperin, Patrick Lee, Brad Ramshaw, George Sawatzky, Mathias Scheurer, Joerg Schmalian, Louis Taillefer, Alexei Tsvelik, and Ya-Hui Zhang for valuable discussions.
This research was supported by the U.S. National Science Foundation grant No. DMR-2245246 and by the Simons Collaboration on Ultra-Quantum Matter which is a grant from the Simons Foundation (651440, S.S.). P.M.B. acknowledges support by the German National Academy of Sciences Leopoldina through Grant No.~LPDS 2023-06 and the Gordon and Betty Moore Foundation’s EPiQS Initiative Grant GBMF8683. M.C. acknowledges funding from Amazon Web Services, AWS Quantum Program National Science Foundation (PHY-2317110). We acknowledge the usage of High Performance Computing resources at the Institute of Mathematical Sciences.

\appendix

\section{Monte Carlo results for the superconducting and charge density wave fluctuations}
\label{sec:KT}

\begin{figure}
    \centering
    \includegraphics[width=0.9\linewidth]{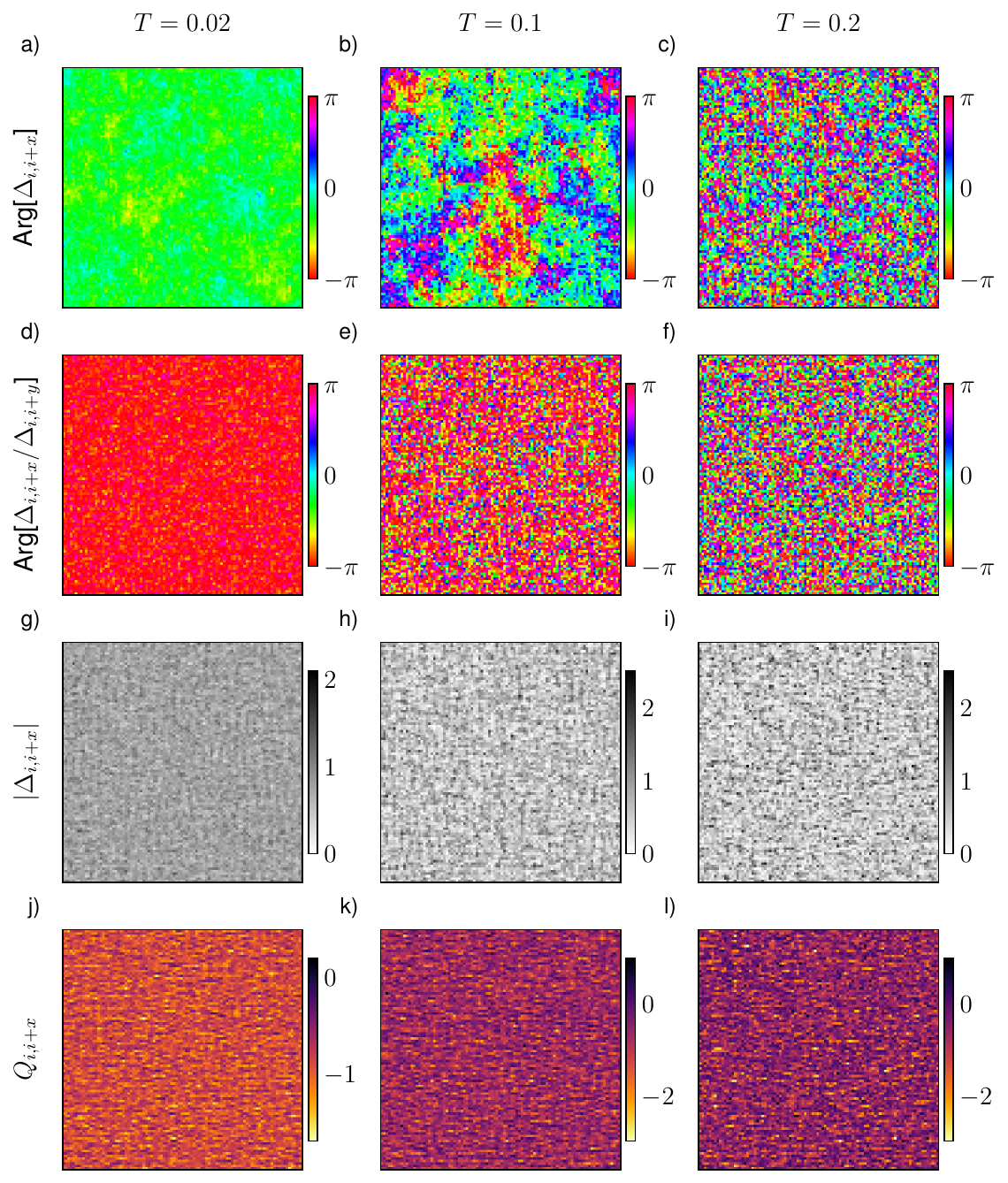}
    \caption{The variation of phase (a-c), relative phase between $x$ and $y$-bonds (d-f) and modulus (g-i) of the bond superconducting order parameter and magnitude of the bond density order parameter (j-l) for three different temperatures across the Kosterlitz-Thouless transition on a $96\times 96$ lattice. The transition temperature for this lattice size is $T_c \simeq 0.09$.}
    \label{fig:MCorders}
\end{figure}

Similar to the plot in Fig.~\ref{fig:BondDensPairingPhLowT}, we show the variation of the magnitude and phase of the superconducting order parameter and the bond order with temperature, across the Kosterlitz-Thouless transition in Fig.~\ref{fig:MCorders}.
Note the appearance of $d$-wave pairing correlations as the temperature is lowered, with no significant correlations in bond order. Only at very low temperatures are we able to discern the correlations between bond order and vortices, as shown in Fig.~\ref{fig:BondDensPairingPhLowT}.


\section{Fermion Hamiltonian and spectrum}
\label{sec:fermions}

The initial discussion of this section will follow a different route from that of ALM summarized in Section~\ref{sec:ALM}. Here we use the structure of the spin liquid to deduce the properties of the chargon $B$, and the connection to the electron operator.

First, we describe the effective Hamiltonian for the spinons $f_{\vi \alpha}$ of the spin liquid of the FL* state. The spinons are connected to the $S=1/2$ spin operator on site $\vi$ via
\begin{align}
    {\bm S}_{\vi} = \frac{1}{2} f_{\vi \alpha}^\dagger {\bm \sigma}_{\alpha\beta}f_{\vi \beta} \,, \label{S2f}
\end{align}
where ${\bm \sigma}$ are the Pauli matrices, and $\alpha,\beta \equiv \uparrow, \downarrow$.

All we need to know about the spin liquid are the projective symmetry transformations of the spinons $f_{\vi \alpha}$, and of their coupling to the lattice SU(2) gauge fields $U_{\vi\vj}$. The projective transformations of $B$ fields then follow from those of the $f_{\vi \alpha}$, as their composite is an electron which does not transform projectively. The symmetry transformations are specified in Table~\ref{tab1}. 
The $f_{\vi\alpha}$ spinons couple minimally to the SU(2) fields when placed in the Nambu form
\begin{align}
\psi_{\vi} \equiv \left( \begin{array}{c} f_{\vi\uparrow} \\ f_{\vi\downarrow}^\dagger \end{array} \right)\,.
\label{eq:Nambu1}
\end{align}
As we review below, these spinon properties {\it fully\/} determine the structure of the effective action for the complex lattice scalar doublet $B_{a\vi }$. 

The analysis is clearest upon introducing a matrix notation for the fermions and bosons \cite{Affleck-SU2,Fradkin88}:
\begin{align}
& \mathcal{C}_\vi \equiv \left(
\begin{array}{cc}
c_{\vi \uparrow} & - c_{\vi \downarrow} \\
c_{\vi \downarrow}^\dagger & c_{\vi \uparrow}^\dagger
\end{array}
\right) \quad , \quad
\mathcal{F}_\vi \equiv \left(
\begin{array}{cc}
f_{\vi \uparrow} & - f_{\vi \downarrow} \\
f_{\vi \downarrow}^\dagger & f_{\vi \uparrow}^\dagger
\end{array}
\right) \quad , \nonumber \\
   &~~~~~~~~~~~~~~~~~ \mathcal{B}_\vi \equiv \left( \begin{array}{cc} B_{1\vi} & - B_{2\vi}^\ast \\ B_{2\vi} & B_{1\vi}^\ast \end{array} \right) \,,
\end{align}
where $c_{\vi \alpha}$ are the electrons in the top layer. These matrices obey the `reality' condition 
\begin{align}
\mathcal{C}_i^\dagger = \sigma^y \mathcal{C}_i^{T} \sigma^y\,, 
\end{align}
and similarly for $\mathcal{F}$ and $\mathcal{B}$.
Then the SU(2) gauge transformation $V_\vi$ acts on the fields as
\begin{align}
\mathcal{C}_\vi \rightarrow \mathcal{C}_\vi \quad & , \quad 
\mathcal{F}_\vi \rightarrow V_\vi \, \mathcal{F}_\vi   \nonumber \\
\mathcal{B}_\vi \rightarrow V_\vi \, \mathcal{B}_\vi \quad & , \quad 
U_{\vi\vj} \rightarrow  V_\vi \, U_{\vi\vj} \, 
V_\vj^{\dagger} \,. \label{eq:gauge}
\end{align}
A global SU(2) spin rotation $\Omega$ on $(c_\uparrow, c_\downarrow)$ acts as
\begin{align}
\mathcal{C}_\vi \rightarrow \mathcal{C}_\vi \, \sigma^z \Omega^T \sigma^z \quad & , \quad 
\mathcal{F}_\vi \rightarrow \mathcal{F}_\vi \, \sigma^z \Omega^T  \sigma^z  \nonumber \\
\mathcal{B}_\vi \rightarrow  \mathcal{B}_\vi \quad & , \quad 
U_{\vi\vj} \rightarrow  U_{\vi\vj}  \,.
\label{eq:spin}
\end{align}
Finally, the U(1) charge conservation symmetry acts as
\begin{align}
\mathcal{C}_\vi \rightarrow  \Theta \,\mathcal{C}_\vi \, \quad & , \quad 
\mathcal{F}_\vi \rightarrow \mathcal{F}_\vi   \nonumber \\
\mathcal{B}_\vi \rightarrow  \mathcal{B}_\vi \, \Theta^\dagger \quad & , \quad 
U_{\vi\vj} \rightarrow  U_{\vi\vj}  \,,
\label{eq:charge}
\end{align}
where 
\begin{align}
    \Theta = \left( \begin{array}{cc}
        e^{i\theta} & 0 \\
         0 & e^{-i \theta}
    \end{array}
    \right)\,.
\end{align}
The gauge, spin rotation, and charge conservation symmetries above are consistent with 
the following operator correspondence between the electrons and the Higgs boson and the spinons 
\begin{align}
    \mathcal{C}_{\vi}^{\vphantom\dagger} \sim \mathcal{B}_\vi^\dagger  \,\mathcal{F}_\vi^{\vphantom\dagger} \,. \label{eq:CBF}
\end{align}
In terms of its matrix components, we can write Eq.~(\ref{eq:CBF}) as 
\begin{align}
c_{\vi \alpha}^{\dagger} \sim
B_{1\vi}^{\vphantom\dagger} f_{\vi \alpha}^\dagger + B_{2\vi}^{\vphantom\dagger} \varepsilon_{\alpha\beta}^{\vphantom\dagger} f_{\vi \beta}^{\vphantom\dagger}  \,, \label{cBf}
\end{align}
where $\varepsilon_{\alpha\beta}$ is the unit antisymmetric tensor for spin SU(2).
From Eq.~(\ref{eq:CBF}) we can also deduce that
\begin{align}
B_\vi \equiv \left( \begin{array}{c} B_{1\vi} \\ B_{2\vi} \end{array} \right)
\end{align}
couples minimally to $U_{ij}$, and that the $B$ fields also experiences the flux in Eq.~(\ref{txty}).

From the gauge transformations in Eq.~(\ref{eq:gauge}), and the global spin rotation in Eq.~(\ref{eq:spin}), we anticipate a spinon hopping term of the form
\begin{align}
    \mbox{Tr} \left( \mathcal{F}_{\vi }^\dagger U_{\vi \vj}^{\phantom\dagger} \mathcal{F}_{\vj }^{\phantom\dagger} \right)
\end{align}
which is invariant under both these transformations. However, the identity
\begin{align}
     \mbox{Tr} \left( \mathcal{F}_{\vi}^\dagger  \mathcal{F}_{\vj }^{\phantom\dagger} \right) = -\mbox{Tr} \left( \mathcal{F}_{\vj }^\dagger  \mathcal{F}_{\vi }^{\phantom\dagger} \right)
\end{align}
requires a pure-imaginary hopping in a Hermitian Hamiltonian in the mean-field approximation with $U_{\vi \vj} = \mathbf{1}$. Hence we have the nearest-neighbor spinon Hamiltonian of the $\pi$-flux spin liquid:
\begin{align}
    \mathcal{H}_{\rm SL} [f] & = -\frac{iJ}{2} \sum_{\langle \vi \vj \rangle} e_{\vi \vj} \left[  \mbox{Tr} \left( \mathcal{F}_{\vi}^\dagger U_{\vi \vj}^{\phantom\dagger} \mathcal{F}_{\vj }^{\phantom\dagger} \right) -  \mbox{Tr} \left( \mathcal{F}_{\vj }^\dagger U_{\vj \vi}^{\phantom\dagger} \mathcal{F}_{\vi }^{\phantom\dagger} \right) \right] \nonumber \\
    & = -i J \sum_{\langle \vi \vj \rangle} e_{\vi \vj} \left[ \psi_{\vi}^\dagger U_{\vi \vj}^{\phantom\dagger} \psi_{\vj}^{\phantom\dagger} - \psi_{\vj}^\dagger U_{\vj \vi}^{\phantom\dagger} \psi_{\vi}^{\phantom\dagger}\right],
    \label{eq:fermionhop}
\end{align}
where $e_{\vi\vj}$ was defined in Eq.~(\ref{su2ansatz}).
The pure imaginary hopping term for the fermions in Eq.~(\ref{eq:fermionhop}) is responsible for the pure imaginary hopping term for the bosons in Eq.~(\ref{Eq:quadraticB}). The dispersion of the fermions at $U_{\vi\vj}=\mathbbm{1}$ is given by the momentum-dependent terms in the boson dispersion Eq.~(\ref{eq:bosondisp}), with $w$ replaced by $J$. However, since the fermions are at half-filling, the most relevant momenta are now near the zero-energy points, $(0,0)$ and $(0, \pi)$. Here, the dispersion is that of massless Dirac fermions, yielding $N_f=2$ fermionic Dirac spinons in the low-energy SU(2) gauge theory.

Next, we describe the effective Hamiltonian for the Fermi surface of electron-like quasiparticles. Here we use the Ancilla Layer Model (ALM): this maps the single-band Hubbard model to a model of free electrons interacting with a bilayer antiferromagnet of spins ${\bm S}_1$ and ${\bm S}_2$; for a review, see Refs.~\cite{Boulder25,ICTP25}. The quasiparticles are realized by a Kondo lattice heavy Fermi liquid of the electrons coupled to the ${\bm S}_1$ spins; the spin liquid of the FL* state is realized by the ${\bm S}_2$ layer, which we present using the $f$ spinons, as in Eq.(\ref{S2f}). Representing the spins ${\bm S}_1$ of the Kondo lattice by spinons $f_{1,\vi \alpha}$ as
\begin{align}
    {\bm S}_{1, \vi} = \frac{1}{2} f_{1,\vi \alpha}^\dagger {\bm \sigma}_{\alpha\beta}f_{1,\vi \beta} \,, \label{S1f}
\end{align}
we use the Hamiltonian of Mascot {\it et al.\/} \cite{Mascot22}
\begin{align}\label{Top2layers_Hamiltonian}
  \mathcal{H}_{\rm KL} [c,f_1] & = \sum_{\vi,\vj} \left[
  t^{c}_{\vi \vj} c_{\vi \alpha}^\dagger c_{\vj \alpha}^{\vphantom\dagger} +  t^{f}_{\vi \vj} f_{1,\vi \alpha}^\dagger f_{1,\vj \alpha}^{\vphantom\dagger}
  \right] \nonumber \\
  & + \sum_{\vi} \left[ \phi\, c_{\alpha}^\dagger f_{1,\vi \alpha}^{\vphantom\dagger} + \mbox{H.c}\right]\,.
\end{align}
This Hamiltonian has the form of a standard Kondo lattice heavy Fermi liquid Hamiltonian of conduction electrons $c_{\vi \alpha}$ with a hybridization of $\phi$ to the localized moments ${\bm S}_{1,i}$ represented by the fermions $f_{1,\vi\alpha}$. 
The field $\phi$ is originally obtained by decoupling the Kondo exchange, but we assume here it is condensed and treat it as a $c$-number; consequently, $\phi$ and $f_{1\alpha}$ do not carry any gauge charges. Then diagonalizing the above Hamiltonian yields a heavy Fermi liquid with 4 hole pockets, each of fractional area $p/8$. The magnitude of $\phi$ determines the pseudogap in the fermion spectrum in the antinodal region near momenta $(\pi, 0)$ and $(0,\pi)$ \cite{ZhangSachdev_ancillaA,Mascot22A}, and its dispersion computed from Eq.~(\ref{Top2layers_Hamiltonian}) agrees with experimental observations \cite{{Shen11A}}; we used the values $t_{\bm{i},\bm{i}+\hat{x}}^c = t_{\bm{i},\bm{i}+\hat{y}}^c =-.22$ eV, $t_{\bm{i},\bm{i}\pm\hat{x}\pm\hat{y}}^c =.034$ eV, $t_{\bm{i},\bm{i}\pm2\hat{x}}^c =t_{\bm{i},\bm{i}\pm 2\hat{y}}^c =-.036$ eV, $t_{\bm{i},\bm{i}\pm\hat{x}\pm2\hat{y}}^c =t_{\bm{i},\bm{i}\pm2\hat{x}\pm\hat{y}}^c =.007$ eV for the $c$ electron hoppings, $t_{\bm{i},\bm{i}+\hat{x}}^{f_1} = t_{\bm{i},\bm{i}+\hat{y}}^{f_1} =.1$ eV, $t_{\bm{i},\bm{i}\pm\hat{x}\pm\hat{y}}^{f_1} =-.03$ eV, and $t_{\bm{i},\bm{i}\pm2\hat{x}}^{f_1} =t_{\bm{i},\bm{i}\pm 2\hat{y}}^{f_1} =-.01$ eV,  for the $f_1$ fermion hoppings while the hybridization $\phi = .36$ eV unless otherwise specified in the text. All Hamiltonian couplings are given in units of eV throughout the text and all temperatures specified are also in units of $eV$ (after taking $k_B=1$). 

Quantum fluctuations of $\phi$ drive the higher temperature crossover with increasing doping from FL* to the Fermi liquid via the strange metal, and this has been studied elsewhere \cite{ZhangSachdev_ancillaA,ZhangSachdev_ancilla2A,Random_ancilla,SSORE,LPS24}. Our focus here is the fate of FL* as we lower the temperature, and thus we can ignore the fluctuations of $\phi$ about its mean field value. Correspondingly, we can also ignore fluctuations of the gauge fields associated with the spinon decomposition in Eq.~(\ref{S1f}), since the gauge field is Higgsed by $\phi$. In this situation, the $f_{1,\alpha}$ fermions can be interpreted as electrons, because they have the same quantum numbers as electrons.

Finally, we couple the $f_{\vi \alpha}$ spinons to the Kondo lattice electrons. This coupling is realized by the bosons $B$, which is the decoupling field of the $J_\perp$ rung-exchange between the ${\bm S}_1$ and ${\bm S}_2$ layers. The gauge and symmetry transformations in Eqs.~(\ref{eq:gauge},\ref{eq:spin}) allow the on-site hybridization associated with Eqs.~(\ref{eq:CBF},\ref{cBf}):
\begin{align}
  & \mathcal{H}_{\rm Y} [c,f_1,f]   = -\frac{1}{2} \sum_\vi \left[ i g_1 \, \mbox{Tr} \left(\mathcal{C}_{\vi}^\dagger \mathcal{B}_{\vi}^\dagger \mathcal{F}_i^{\vphantom\dagger}\right) \right. \nonumber \\
  &~~~~~~~~~~~~~~ \left. + i g_2 \, \mbox{Tr} \left(\mathcal{F}_{1,\vi}^\dagger \mathcal{B}_{\vi}^\dagger \mathcal{F}_i^{\vphantom\dagger} \right) + \mbox{H.c.} \right] \nonumber \\
   & =  \sum_\vi \left[ i g_1 \left( B_{1\vi}^{\vphantom\dagger} f_{\vi\alpha}^\dagger c_{\vi \alpha}^{\vphantom\dagger} - B_{2 \vi}^{\vphantom\dagger} \varepsilon_{\alpha\beta}^{\vphantom\dagger} f_{\vi \alpha}^{\vphantom\dagger} c_{\vi \beta}^{\vphantom\dagger} \right)
   + \mbox{H.c.} \right. \nonumber \\
   & ~+ \left. i g_2 \left( B_{1\vi}^{\vphantom\dagger} f_{\vi\alpha}^\dagger f_{1,\vi \alpha}^{\vphantom\dagger} - B_{2 \vi}^{\vphantom\dagger} \varepsilon_{\alpha\beta}^{\vphantom\dagger} f_{\vi \alpha}^{\vphantom\dagger} f_{1,\vi \beta}^{\vphantom\dagger} \right)
   + \mbox{H.c.}  \right],
   \label{Yukawa}
\end{align}
where
\begin{align}
    \mathcal{F}_{1,\vi} = \left(
\begin{array}{cc}
f_{1,\vi \uparrow} & - f_{1,\vi \downarrow} \\
f_{1,\vi \downarrow}^\dagger & f_{1,\vi \uparrow}^\dagger
\end{array}
\right)\,.
\end{align}
We have introduced two Yukawa couplings $g_1$, $g_2$. The implicit temperature dependence in the Higgs potential $\mathcal{E}_2+\mathcal{E}_4$ can be transferred by a rescaling of $B$ into a temperature dependence of the coupling $g_{1,2}$.

For the convenience of the readers, we summarize the complete ALM Hamiltonian used in this work for the bosons $B$, $U$, and the fermions of the three layers $c$, $f_1$, $f$:
\begin{align}
\mathcal{H}_{\rm ALM} = & \mathcal{H}_{\rm KL}[c,f_1] + \mathcal{H}_{\rm SL}[f] + \mathcal{H}_{\rm Y} [c,f_1,f] \nonumber \\ &+ \mathcal{E}_2 [B,U] + \mathcal{E}_4 [B,U]
\label{eq:complete}    
\end{align}
which are specified in Eqs.~(\ref{Top2layers_Hamiltonian}), (\ref{eq:fermionhop}), (\ref{Yukawa}), (\ref{Eq:quadraticB}), and (\ref{Eq:quarticB}), denoting the Kondo lattice (KL), the spin liquid (SL), the Yukawa coupling between them (Y), 
and the quadratic and quartic energy functionals for the bosons respectively. 

\section{Gaussian Sampling}
\label{sec:Gaussianfermion}

Our interest in the fermion spectra is primarily in the normal state. Instead of the expensive Monte Carlo simulations in the main article, this supplement discusses a simpler gaussian approximation for the fluctuations of $B_\vi$, which we can expect to be a reasonable approximation in the high-temperature phase. The gaussian approximation is equivalent to approximating the energy by $\mathcal{E}_2$ in Eq.~(\ref{Eq:quadraticB}), and suppressing the gauge fields by setting $\kappa=\infty$.  Given that pseudogap Fermi surfaces without thermal fluctuations enclose an area $\frac{p}{8}$ where $p$ is generally expected to be small within the pseudogap phase, a calculation of quantum oscillations requires a system size on the order of a hundred lattice sites for each value of magnetic field and is less practical to study within the fully interacting theory. However, as we will see in Section~\ref{sec:QO}, we can compute quantum oscillations in the presence of thermal fluctuations with qualitatively similar spectral functions within the gaussian theory.
Care must be taken in applying the gaussian approximation to the physical system: in particular, we should view the $T$-dependence of the physical quantities as arising not only from the explicit $T$ present in the partition function defined in Eq.~(\ref{Eq:quadraticB}), but also from a $T$-dependent renormalization of the `mass' $r$ from all the non-gaussian terms.
\begin{figure}
    \centering
    \includegraphics[width=1.1\linewidth]{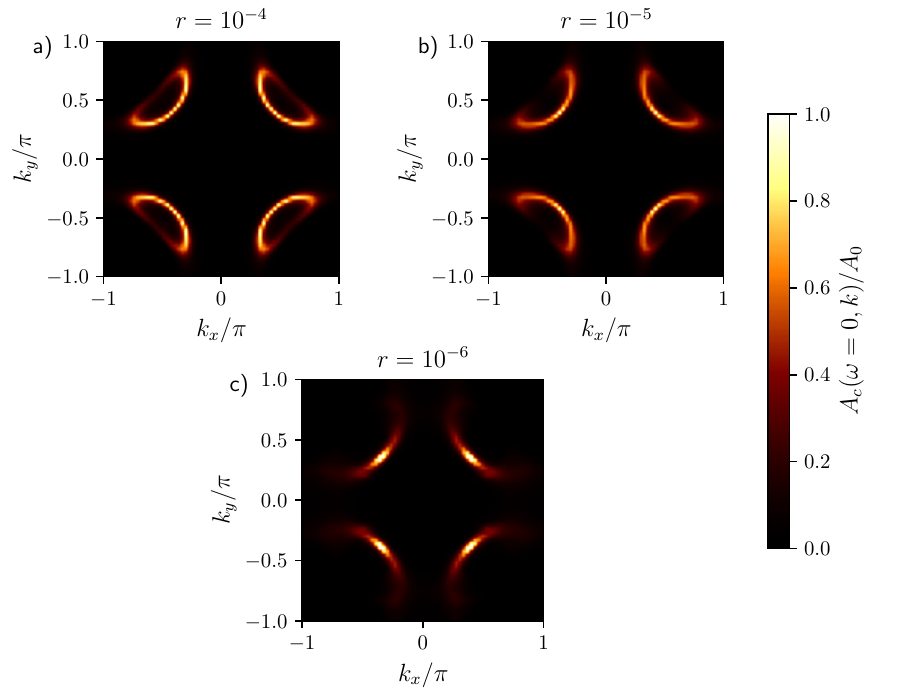}
    \caption{The spectral function in momentum space of the electrons when $B_\vi$ are sampled according only to the gaussian contributions to the quadratic free energy in Eq.~(\ref{Eq:quadraticB}) about the saddle point $U_{\vi\vj}=1$ with $w=0.2/(2\sqrt{2})$ at a temperature $T=2.5\times10^{-5}$ for $r=10^{-4}$ (a), $r=10^{-5}$ (b), $r=10^{-6}$ (c), by averaging over 100 samples each with a broadening parameter $\eta=0.01$.  We use a spin liquid hopping strength $J=0.2/\sqrt{2}$. All other model parameters used are the same as Fig.~\ref{fig:fermionMC}. The spectral intensity is plotted normalized by its maximum value $A_0$ for each value of $r$. }
    
    \label{fig:GaussianBosons_r}
\end{figure}

We perform the sampling over the fields $B_{\bm{k}}$ in the momentum space and compute the real space electronic spectral 
function using the mean-field Hamiltonian of Eqs.~(\ref{eq:fermionhop}),~(\ref{Top2layers_Hamiltonian}), and 
~(\ref{Yukawa}). As in \cite{Christos24}, we set $g_1=0$, $g_2=1$, and use the same hopping parameters as 
\cite{Mascot22}, such that the electrons are at filling $1+p$ with a hole-doping $p=0.2$ when $B=0$. We take a 
value $\phi=0.36$ for the boson coupling in the first two layers in Eq.~(\ref{Top2layers_Hamiltonian}), a spin liquid 
hopping $J=0.2/\sqrt{2}$ and set the boson hopping to $w={J}/{2}$. The resulting momentum space spectral functions 
computed after averaging over 100 samples of $B$ are shown in Fig.~\ref{fig:GaussianBosons_r}(a-c) for different values 
of the boson chemical potential $r$ relative to hopping $w$. In general, we choose a small $r$ (and correspondingly a 
low temperature $T$) such that the correlation length will be large within the gaussian theory.

\begin{figure}
    \centering
    \includegraphics[width=\linewidth]{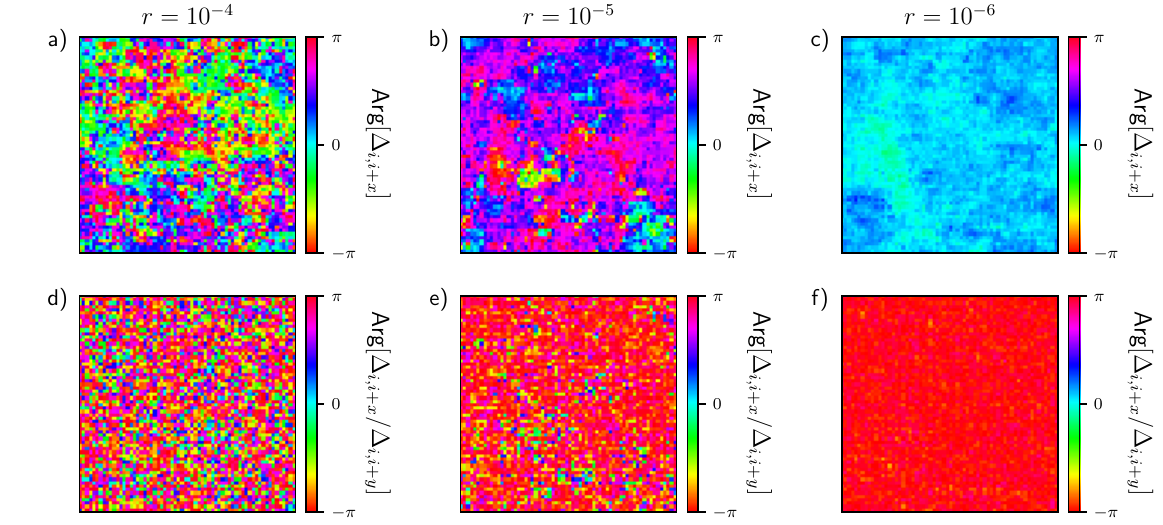}
    \caption{A representative sample of the phase of the bond superconducting order parameter in Eq.~(\ref{Eq:OPs}) for $B_\vi$ sampled according only to the gaussian contributions to the quadratic free energy Eq.~(\ref{Eq:quadraticB}) about the saddle point $U_{\vi\vj}=1$ with $w=0.2/(2\sqrt{2})$ and at a temperature $T =2.5\times10^{-5}$ for $r=10^{-4}$ (a), $r=10^{-5}$ (b), $r=10^{-6}$ (c). We also show the relative phase between the $x$-bond superconducting order parameter and $y$-bond superconducting order parameter (d-f) for the same values of $r$ as in (a-c)}
    \label{fig:GaussianBosons_phase}
\end{figure}

\begin{figure}
    \centering
    \includegraphics[width=\linewidth]{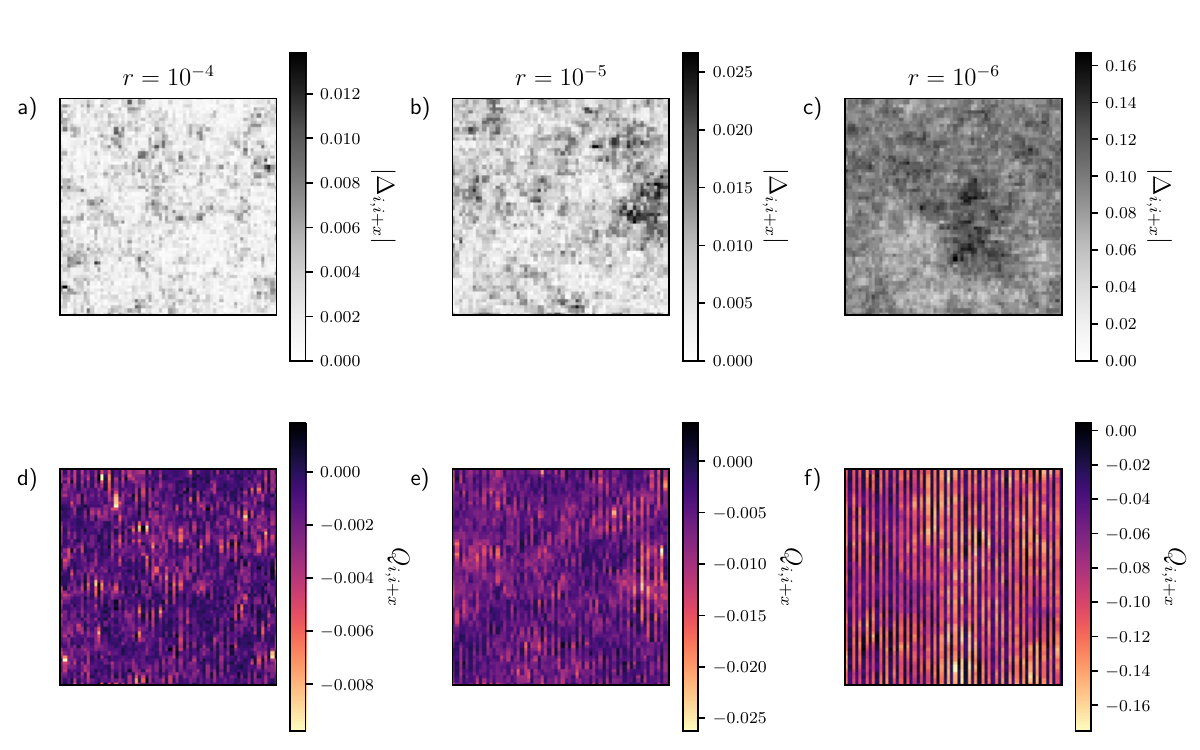}
    \caption{The absolute value of the bond superconducting order parameter in Eq.~(\ref{Eq:OPs}) for a single sample of $B_{\vi}$ sampled according only to the gaussian contributions to the quadratic free energy Eq.~(\ref{Eq:quadraticB}) about the saddle point $U_{ij}=1$ with $w=0.2/(2\sqrt{2})$ and $T =2.5\times10^{-5}$ (same as Fig.~\ref{fig:GaussianBosons_phase}) for $r=10^{-4}$ (a), $r=10^{-5}$ (b), $r=10^{-6}$ (c). We also show the value of the bond density order parameter (d-f) for the same values of $r$ as in (a-c).}
    \label{fig:GaussianAbs}
\end{figure}

We show that as $r$ decreases, resulting in the presence of long-range correlations, the electronic spectral intensity associated with the backside of the small hole pockets is diminished. A representative sample of the phase of one of the bond superconducting order parameters is shown in Fig.~\ref{fig:GaussianBosons_phase}(a-f) for different values of $r$, along with the absolute value of the bond superconducting and bond density order parameters in Fig.~\ref{fig:GaussianAbs}(a-f). As $r$ decreases, patches of $d$-wave phase coherence form with increasing correlation length, co-existing with the patches of period 2 stripe order (which is degenerate with $d$-wave superconductivity at the level of $\mathcal{E}_2$) as indicated by the modulations of $Q_{ij}$ in Fig.~\ref{fig:GaussianAbs}(d-f). The evolution of the Fermi surface for a fixed value of $r$ and varying $\beta$ is shown in Fig.~\ref{fig:Photo_Osc}a. At the level of Eq.~(\ref{Eq:quadraticB}), varying $\beta$ rescales the variance of $B$, resulting in the diminishing of the backside pocket spectral intensity as temperatures are increased and a spectral function resembling the Fermi arcs 
observed in photoemission experiments.

It is clear from the results in the present section that with gaussian fluctuations of $B$ alone, the electronic spectra are sharper and less diffused than those obtained including $B$ and $U$ fields from full Monte Carlo simulations in the main article. We attribute this difference to the fluctuations of the SU(2) gauge fields $U$, which are not included here. Nevertheless, even in the gaussian theory, we see evidence of the formation of Fermi arcs.


\subsection{Quantum Oscillations}
\label{sec:QO}

Finally, we study the electronic Fermi surface in the presence of gaussian fluctuations by computing oscillations of the density of states induced by a U(1) magnetic field. To account for a non-zero magnetic field associated with the U(1) charge in Eq.~(\ref{eq:charge}), we introduce a Peierls phase $t_{\vi\vj}\rightarrow t_{\vi\vj} \exp \left( i\frac{e}{\hbar}\int_{\vi}^{\vj} \bm{A}\cdot d\bm{s}\right)$ to all hoppings in the first and second layers of Eq.~(\ref{Top2layers_Hamiltonian}) as well as to the boson hopping $w$ as defined in Eq.~(\ref{BAhop}). The fermions moving in $\pi$-flux are still deconfined and are not coupled to the U(1) gauge field. In our calculations, we choose a Landau gauge such that the mean-field Hamiltonian is translationally invariant in the $y$-direction and a system size of $800\times800$. We compute the density of states at the Fermi level of the $c$ electron for the varying magnetic field $H$. Using the kernel polynomial method \cite{Wei_e_2006} to compute the density of states at the Fermi level and taking an expansion up to $N=4800$ polynomials, the trace over real space with $M=4$ real space vectors is computed stochastically. 

\begin{figure}
    \centering
    \includegraphics[width=4in]{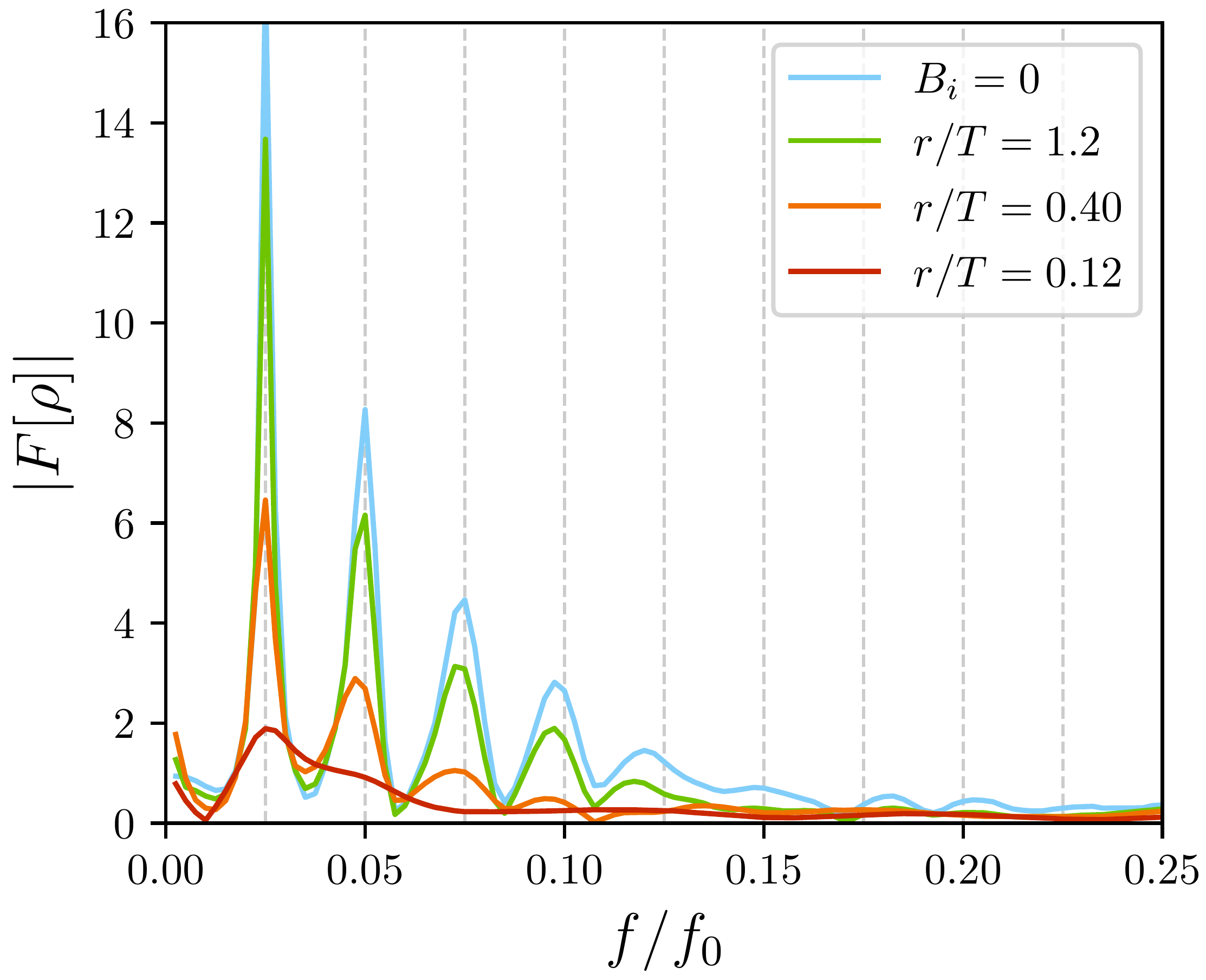}
    \caption{Fourier transform of the density of states, $\rho$, in Fig.~\ref{fig:Photo_Osc}b as a function of frequency normalized by $f_0$, the frequency corresponding to the total Brillouin zone area. Black dashed lines denote the normalized frequency ${f}/{f_0}=n {p}/{8}$ for $n=1,2,3,...$, corresponding to the expected frequency of small hole pockets with associated area $p/8$. The oscillations were computed averaged over 3 samples per value of $T$.}
    \label{fig:GaussianQO}
\end{figure}

The density of states as a function of ${1}/{H}$ is shown in Fig.~\ref{fig:Photo_Osc}b, and its Fourier transform is shown in  Fig.~\ref{fig:GaussianQO}. 
Using the same values of $T$ and $r$ as in Fig.~\ref{fig:Photo_Osc}a, we find oscillations of the density of states with a periodicity corresponding to a frequency $f=({p}/{8})f_{0}=0.025f_{0}$ for hole-doping $p=0.2$ where $f_{0}={h}/({ea_0^2})$ is the frequency corresponding to the area of the Brillouin zone. 


\section{One-loop self-energy in the ancilla layer theory with quantum bosons}
\label{app:oneloop}
\begin{figure}
    \centering
    \includegraphics[width=0.25\linewidth]{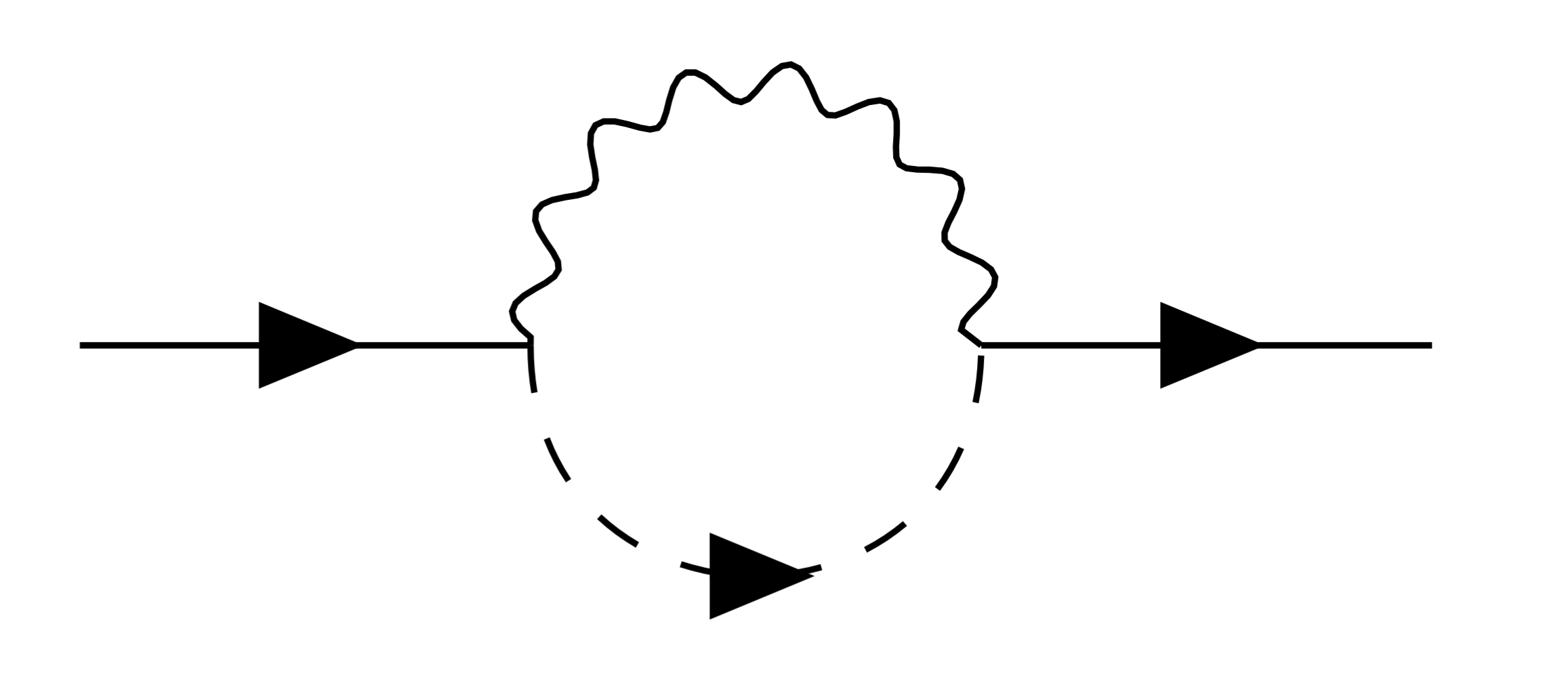}
    \caption{Feynman diagram for the 1-loop self-energy within the ancilla layer theory. The solid lines represent incoming and outgoing $c$ or $f_1$ electrons, the dashed line the propagator of the fermionic spinon $G$, and the wiggly one the propagator of the bosonic chargon $D$.}
    \label{fig:Fock_self_energy_diagram}
\end{figure}

In this section, we present a calculation of the electron spectral function up to second order in the electron-spinon-chargon coupling, while including quantum fluctuations of $B$. The results here for the electron spectral functions are fairly similar to those obtained by the Monte Carlo simulations, and the gaussian approximation of Appendix~\ref{sec:Gaussianfermion}.

We consider the Hamiltonian given by Eqs.~\eqref{eq:fermionhop}, \eqref{Top2layers_Hamiltonian}, \eqref{Yukawa} and \eqref{Eq:quadraticB}, where we only focus on the quadratic part of Eq.~\eqref{Eq:quadraticB} and set $U_{\vi\vj}=\mathbbm{1}$. The second-order correction to the mean-field propagator can be represented by the Feynman diagram in Fig.~\ref{fig:Fock_self_energy_diagram}. Evaluating the diagram, we obtain the self-energy (in real space and imaginary time) at finite temperature $T$:
\begin{align}\label{eq: self energy}
    \Sigma_{\vi \vj}(\tau,\tau')= & \left(
    \begin{array}{cc}
        \Sigma^{cc} & \Sigma^{cf}  \\
        \Sigma^{fc} & \Sigma^{ff}
    \end{array}\right) = \left(
    \begin{array}{cc}
        g_1^2 & g_1\,g_2  \\
        g_1 \,g_2 & g_2^2
    \end{array}\right)
    \nonumber \\
   & ~~~~~~\times \mathrm{Tr}\left[G_{\vi \vj}(\tau,\tau')\,D_{\vj \vi}(\tau',\tau)\right]\,,
\end{align}
where the couplings $g_{1,2}$ have been defined in Eq.~\eqref{Yukawa}.
We have also defined the spinon propagator $G$ as 
\begin{equation}
    G_{\vi \vj}(\tau,\tau') = T\!\sum_{n=-\infty}^{+\infty}\,e^{i\omega_n(\tau-\tau')}\,\left[i\omega_n\mathbbm{1}\delta_{\vi \vj}+\frac{J}{\sqrt{2}}\chi_{\vi \vj}\right]^{-1}\,,
\end{equation}
and the chargon propagator as
\begin{align}
   &  D_{\vi \vj}(\tau,\tau') =  T\!\sum_{n=-\infty}^{+\infty}\,e^{i\Omega_n(\tau-\tau')} \nonumber \\
    &~~~ \times \left[ (-\Omega_n^2-r-2\sqrt{2}w)\mathbbm{1}\delta_{\vi \vj}+\frac{w}{\sqrt{2}}\chi_{\vi \vj}\right]^{-1}\,, 
\end{align}
where $\omega_n=(2n+1)\pi T$ and $\Omega_n=2n\pi T$ are the fermionic and bosonic Matsubara frequencies, respectively. Here $\chi_{\vi \vj}$ are the $\pi$-flux hoppings expressed in the so-called $d$-wave gauge (the analogue of $i\, e_{\vi \vj}$ of the main text, which however uses a different gauge),
\begin{equation}
    \chi_{\vi \vj}=\begin{cases}
            \tau^3+\tau^1\quad\text{if}\quad \vj=\vi\pm\hat{\boldsymbol{x}}\,,\\
            \tau^3-\tau^1\quad\text{if}\quad \vj=\vi\pm\hat{\boldsymbol{y}}\,,\\
            0\quad\text{otherwise,}
        \end{cases}
\end{equation}
with $\tau^a$ being the Pauli matrices.
Note that the self-energy in Eq.~\eqref{eq: self energy} is SU(2) gauge invariant. 

Defining $\Sigma^{(0)}(\bk,\omega)$ as the Fourier transform of $\mathrm{Tr}\left[G_{\vi \vj}(\tau,\tau')\,D_{\vj \vi}(\tau',\tau)\right]$ analytically continued to real frequencies, we obtain
\begin{equation}
    \begin{split}
        \Sigma^{(0)}(\mathbf{k}, \omega) &= \int_{\mathbf{q}} \sum_{b, f = \pm} \frac{\mathrm{Tr}\left[U_\bq^b\, U^f_{\bk+\bq}\right]}{4 \mathcal{E}_{b, \mathbf{q}}}\times  \\
        \Bigg\{&\left[ T_{f,\bk+\bq} + C_{b,\bq} \right] \frac{1}{ - \omega_+ + \mathcal{E}_{b, \mathbf{q}} + E_{f, \mathbf{k} + \mathbf{q}}} \\
        + &\left[ T_{f,\bk+\bq} - C_{b,\bq} \right] \frac{1}{\omega_+ + \mathcal{E}_{b, \mathbf{q}} - E_{f, \mathbf{k} + \mathbf{q}}} 
        \Bigg\}\,,        
    \end{split}
\end{equation}
where $\omega_+=\omega+i0^+$ and $E_{\pm,\bk}=\pm 2J\sqrt{\cos^2 k_x+\cos^2 k_y}$, $\mathcal{E}_{\pm,\bq}=\sqrt{r+2\sqrt{2}w\pm 2w\sqrt{\cos^2 q_x+\cos^2 q_y} }$, $T_{f,\bk}=\tanh\left( \frac{E_{f, \mathbf{k}}}{2T} \right)$, and $C_{b,\bq}=\coth\left( \frac{\mathcal{E}_{b, \mathbf{q}}}{2T} \right)$. Note that $r\geq 0$ since the bosonic excitation energies must be real. The coherence matrices are defined as 
\begin{subequations}
    \begin{align}
        &U^+_\bk =\left(
        \begin{array}{cc}
            v_\bk^2 & -u_\bk v_\bk  \\
            -u_\bk v_\bk & u_\bk^2
        \end{array}
        \right)\,,\\
        &U^-_\bk =\left(
        \begin{array}{cc}
            u_\bk^2 & u_\bk v_\bk  \\
            u_\bk v_\bk & v_\bk^2
        \end{array}
        \right)\,,\\
        &u_\bk=\sqrt{\frac{1}{2}\left(1+\frac{\cos k_x+\cos k_y}{\sqrt{2}\sqrt{\cos^2 k_x + \cos^2 k_y}}\right)}\,,\\
        &v_\bk=\sqrt{\frac{1}{2}\left(1-\frac{\cos k_x+\cos k_y}{\sqrt{2}\sqrt{\cos^2 k_x + \cos^2 k_y}}\right)}\,.
    \end{align}
\end{subequations}
Finally, the electron Green's function can be calculated as 
\begin{align}
    & \hat{G}(\bk,\omega)=\left[\omega_+\mathbbm{1}-\left(
    \begin{array}{cc}
        \epsilon^c_\bk & \phi \\
        \phi^* & \epsilon^f_\bk
    \end{array}
    \right) \right. \nonumber \\
    &~~~~~~~~~~~~~ \left. - \left(
    \begin{array}{cc}
        g_1^2 & g_1\,g_2  \\
        g_1 \,g_2 & g_2^2
    \end{array}\right)\Sigma^{(0)}(\bk,\omega)
    \right]^{-1}\,,
\end{align}
with $\epsilon^c_\bk$ and $\epsilon^f_\bk$ the Fourier transforms of $t^c_{\vi \vj}$ and $t^f_{\vi \vj}$ in Eq.~\eqref{Top2layers_Hamiltonian}. The electron spectral function can be extracted from the Green's function, 
\begin{equation}
    A(\bk,\omega)=-\frac{1}{\pi}\mathrm{Im}\,\hat{G}_{cc}(\bk,\omega)\,.
\end{equation}

\begin{figure}
    \centering
    \includegraphics[width=1.0\linewidth]{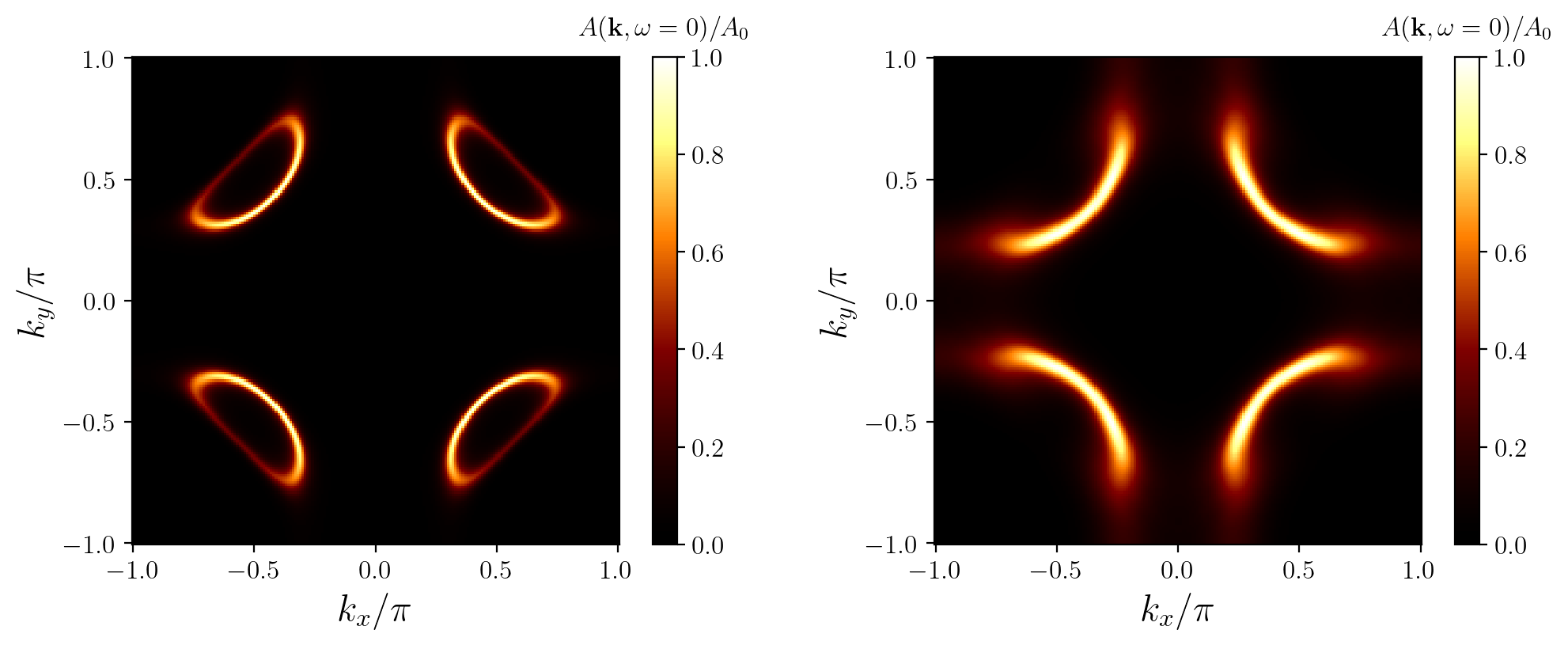}
    \caption{Zero frequency cuts of the electron spectral function. Left: mean-field result ($\Sigma^{(0)}(\bk,\omega)=0$). Right: Fock self-energy included. For this calculation, we used a broadening of the electron Green's function equal to 0.01. The spectral intensity is plotted normalized by its maximum value $A_0$.}
    \label{fig:Fock_self_energy}
\end{figure}

In Fig.~\ref{fig:Fock_self_energy} we show the comparison between the mean-field electron spectral function, i.e., with $\Sigma^{(0)}(\bk,\omega)=0$ and the one with $\Sigma^{(0)}(\bk,\omega)$ included. The parameters defining $\epsilon^c_\bk$ and $\epsilon^f_\bk$, as well as the value of $\phi$ are the same as in Figs.~\ref{fig:Photo_before_after} (left), \ref{fig:fermionMC}(a-b), \ref{fig:GaussianBosons_r} and \ref{fig:Photo_Osc}a while the other parameters are $r=0.005$, $w=0.2/(2\sqrt{2})$ (eV) and $J=0.2/\sqrt{2}$, $T=0.1$, $g_1=0$, $g_2=0.15$. We observe that the inclusion of the self-energy provides a decay channel for the electrons that is kinematically more efficient near the backsides of the hole pockets, as the spinon low-energy excitations lie near those, thus washing them out. The inner sides of the pockets, instead, are barely modified by the self-energy inclusion, making the spectral function appear ``arc-like''.

\bibliography{references}


\end{document}